\documentstyle[11pt,aasms4]{article}
\input tcilatex.tex
\received{March 2000 }
\accepted{June 2000 to appear ApJ, 543, November 10}
\begin{document}
\newcommand{\Grad}[1]{\mbox{$\nabla #1$}}
\newcommand{\Div} [1]{\mbox{$\nabla\hspace{-3pt}\cdot\hspace{-2pt}#1$}}
\newcommand{\Curl}[1]{\mbox{$\nabla\hspace{-3pt}\times\hspace{-2pt}#1$}}
\newcommand{\Lap} [1]{\mbox{$\nabla^{2} #1$}}
\title{The Stability of Radiatively Cooled Jets in Three Dimensions}

\author{Jianjun Xu\altaffilmark{1}}
\affil{Department of Astronomy,  The University of Maryland, \\
College Park, Maryland, MD 20742}

\author{Philip E. Hardee}  
\affil{Department of Physics \& Astronomy, The University of Alabama, \\
Tuscaloosa, AL 35487;  hardee@athena.astr.ua.edu}

\author{and \\ James M. Stone}
\affil{Department of Astronomy,  The University of Maryland, \\
College Park, Maryland, MD 20742; jstone@astro.umd.edu}

\altaffiltext{1}{IMS, MCIWorldcom, Vienna, VA 22182}

\baselineskip 11 pt

\begin{abstract}

The effect of optically thin radiative cooling on the Kelvin-Helmholtz
instability of three dimensional jets is investigated via linear
stability theory and nonlinear hydrodynamical simulation.  Two
different cooling functions are considered: radiative cooling is found
to have a significant effect on the stability of the jet in each
case.   The wavelengths and growth rates of unstable modes in the
numerical simulations are found to be in good agreement with
theoretical predictions.  Disruption of the jet is found to be
sensitive to the precessional frequency at the origin with lower
frequencies leading to more rapid disruption.  Strong nonlinear effects
are observed as the result of the large number of normal modes in three
dimensions which provide rich mode-mode interactions.  These mode-mode
interactions provide new mechanisms for the formation of knots in the
flows.  Significant structural features found in the numerical
simulations appear similar to structures observed on protostellar
jets.

\end{abstract}

\keywords{galaxies: jets --- hydrodynamics --- instabilities --- ISM: jets and outflows}

\baselineskip 12 pt

\section{Introduction}

High resolution imaging surveys have revealed the association between
Herbig-Haro (HH) objects and  protostellar jets (\cite{rbglz86};
\cite{sswmw86}; \cite{mbb87}; \cite{ray87}).  Although individual HH
objects can be formed via different mechanisms, most HH objects appear
as bright knots along collimated jets, or associated with the working
surface at the head of the jet, e.g., HH~46/47 (\cite{ge83}), HH~40
(\cite{mbfnss84}), HH~34 (\cite{bmr88}), and HH~111
(\cite{reipurth97}).  Recent HST observations of HH~111
(\cite{reipurth97}) indicate that the bright knots along the jet beam
arise from ``bow'' shocks associated with dense knots inside a
collimated outflow. Sinusoidal distortion of the jet in HH~111 and
staggering of the ``bow'' shocks suggest that the jet beam is helically
twisted.  HST observations of HH~30 (\cite{betal96}) also reveal bright
knots inside the jet beam. The proper motion of these knots has been
detected and the observed bending of the jet changes with time, again
suggesting helical twisting of the jet beam.

A supersonic astrophysical jet is Kelvin-Helmholtz (K-H) unstable [see
Birkinshaw (1991) for a review], and complex structures can be formed
through the growth of instabilities in the jet beam.  The study of the
stability of supersonic astrophysical jets has focused on two
fundamental questions: (1) What is required to make jets sufficiently
stable to propagate large distances (up to a pc in the case of
protostellar jets and hundreds of Kpc in extragalactic sources), and
(2) what emission features, e.g., individual HH objects, are formed
through instabilities in the jet beam?  In protostellar jets, optically
thin radiative cooling can strongly affect the dynamics, e.g., Blondin
et al.\ (1990), Stone \& Norman (1993a, 1993b, 1994), de Gouveia dal
Pino \& Benz (1993).  These earliest numerical studies focused on
propagation of the jets and interactions at the jet front, and showed
that radiative cooling yielded strong fragmentation in a dense shell of
material swept up by the working surface at the jet head.

Interactions behind the jet front and the development of knots in a jet
beam were first considered by Falle et al.\ (1987) who proposed that
standing shocks could form the knots. Subsequently the knots were
observed to have large proper motions (e.g. Eisl\"{o}ffel \& Mundt
1995), and to have a close spacing inconsistent with standing shocks in
the high Mach number flows. Later work showed that pulsation (temporal
variation) of the outflow at the 10\% level could produce aligned knots
with bow shock morphology, e.g., Reipurth et al.\ (1986), Raga \&
Kofman (1992), Stone \& Norman (1993b) de Gouveia Dal Pino \& Benz
(1994); Biro \& Raga (1994); Biro (1996); Suttner et al.\ (1997). It
was also suggested that knots can be the result of K-H flow
instability, e.g., B\"{u}hrke, Mundt, \& Ray (1988), Ray \& Mundt
(1993), Bodo et al.\ (1994), Rossi et al.\ (1997); Downes \& Ray
(1998).  Additionally, the K-H instability could produce helically
twisted structural features.  That asymmetric structural features could
be produced by K-H instability is suggested by recent 2D linear
stability analysis and 2D numerical simulations (\cite{hs97};
\cite{sxh97}).  This work showed that radiative cooling can increase
the growth rate of K-H unstable modes significantly, and revealed a
variety of asymmetric structural features formed as a result of
instability.

In this paper we study the K-H instability of 3D radiatively cooling
jets through a systematic comparison between numerical simulations and
linear stability theory over a wide parameter range.  In \S 2 we
describe the protostellar environment and radiative cooling assumed in
this study.  In \S 3 the 3D linear stability theory is summarized, and
representative normal mode solutions appropriate to the numerical
simulations are obtained.  The numerical simulation results are shown
in \S 4. In \S 4.2 we consider development of the surface wave
modes and in \S 4.3 we consider development of the body wave modes and
wave-wave interaction, and the pressure and velocity fluctuations
arising from wave-wave interactions are compared to computations made
from the linear theory. Our numerical results are summarized in \S 5
and in \S 6 structures  seen in the simulated jets are compared with
observed protostellar jet structures.

\vspace{-0.6cm}

\section{The Protostellar Jet Environment}

Protostellar jets show the same type of spectrum as low excitation HH
objects, and the dominant emission lines include [S~II] and the Balmer
lines. The spectra are best explained as originating from a radiative
shock, with velocity in the range 30 -- 60~km s$^{-1}$ (\cite{ray97}).
Although there are some YSO jets that contain high excitation emission,
this emission can usually be identified with strong bow shocks where
portions of the jet are moving into the ambient medium with a velocity
of a few hundreds of km~s$^{-1}$, comparable to the typical velocities
of $\sim 200$~km~s$^{-1}$ exhibited by the fastest moving knots in the jet
beam. The emission line spectra observed in typical YSO jets indicates
that temperatures in the radiatively cooling regions are $3 \times 10^4
- 10^5$~K, and line ratios, e.g., [S~II] $\lambda =$ 6716{\AA} \&
6731{\AA}, suggest that electron densities are N$_e \sim
100-10,000$~cm$^{-3}$ in the radiatively cooling regions.

In addition to observational evidence that protostellar jets are
overdense with respect to their surroundings (e.g., \cite{ray97}), results from
numerical studies have shown that
a propagating jet creates a cocoon behind the bow shock filled with hot
gas that is less dense than the jet.   Dense molecular outflows which
are often observed to be associated with protostellar jets
(\cite{rc95}) may surround the cocoon.  The kinematic model of a dense
jet embedded in a less dense cocoon which is in turn ensheathed by a
very dense and cold molecular gas is consistent with detailed
observation of a number of systems, e.g., HH~111 whose jet  is
surrounded by a bubble within the larger molecular outflow
(\cite{nvso97}; \cite{crb96}).  Since we focus our study on the
stability properties of the jet beam which is only influenced by the
jet's immediate surroundings, we can model a protostellar jet as a
dense jet beam embedded in a hotter and less dense cocoon.  In
particular, we will assume that the protostellar jet is surrounded by a
uniform, optically thin medium with low density (typically $\sim 100$
cm$^{-3}$) and high temperature (typically $\gtrsim 10^4$ K).  The fact
that protostellar jets remain well collimated for lengths which are
much larger than their radii argues they must be in pressure
equilibrium with their surroundings.  Numerical studies
also have shown that the jet beam remains
cold, neutral, and in approximate pressure balance with the cocoon
medium. Pressure balance implies a jet temperature of $T_{jt} =
(n_{ex}/n_{jt}) T_{ex}$, where the subscripts $ex$ and $jt$ denote the
external cocoon and jet gas respectively.  Observations of some HH jets
(\cite{bmr88}) indicate that typically the internal jet density exceeds
the cocoon density by at least a factor of 10. For $T_{ex} \sim 10^4$~K
and $n_{jt}/n_{ex} = 10$, $T_{jt} \sim 10^3$~K.  This indicates that
the linear analysis of the equilibrium jet can focus on cooling
processes which are effective with jet temperatures $\sim 10^3$~K and
cocoon temperatures $\gtrsim 10^4$~K. Because shocks with $v_s \sim
100$~km~s$^{-1}$ give temperatures as high as $10^6$~K, temperatures
higher than $10^6$~K are considered in the cooling function in the
numerical simulations.

Optical emission from a protostellar jet and from the external gas
results in a loss of internal energy from the system, and the loss of
internal energy through radiative cooling can change the jet dynamics
substantially.  Thus, an important aspect of the numerical simulation
of radiatively cooling jets is an accurate treatment of the
microphysical heating and cooling rates.  While it has been shown that
a non-equilibrium ionization formalism significantly improves the
accuracy of cooling terms (Stone \& Norman 1993a),
incorporating a time-dependent ionization fraction into a linear
stability analysis is intractable.  Thus, the cooling rate is assumed
to be given by equilibrium cooling curves.  In the work to follow two
separate cooling curves are adopted:  (1) the cooling curve for
interstellar gas appropriate to protostellar jets calculated by
Dalgarno \& McCray (1972, hereafter DM), or (2) the curve described by
MacDonald \& Bailey (1981, hereafter MB) for photoionized gas of
reduced metallicity such as that found in elliptical galaxies and
appropriate to extragalactic jets.  While not strictly applicable to
protostellar jets, the MB cooling curve allows us to explore the effect
of temperature dependence of the cooling curve on jet dynamics. The
full DM and MB cooling curves used in the simulations along with the
piecewise power-law fits used in the linear analysis can be found in
Figure 1 of Hardee \& Stone (1997).

\vspace{-0.6cm}

\section{The Linear Theory}

\vspace{-0.2cm}

\subsection{Normal Mode Analysis}

The stability of an adiabatic 3D jet with ``top hat'' profile residing
in a uniform medium has been thoroughly investigated in the literature
(see \cite{birkinshaw91}).  Modifications to the adiabatic 3D theory to
account for radiative cooling are identical to those found in the 2D
theory (\cite{hs97}) and we sketch the results here.  The hydrodynamic
equations of continuity, momentum, and energy are linearized within the
jet and the external (cocoon) gas where ${\bf u} = 0$ and the flow
velocity is reintroduced when solutions are matched at the jet
boundary. The linearized hydrodynamical equations relevant to our model
become
\begin{equation}\label{1}
\frac{\partial \rho_1}{\partial t} + \mbox{\boldmath $\nabla$} \cdot (
\rho_0 \mbox{\boldmath $v_1$ })  =  0 ,
\end{equation}
\begin{equation}\label{2}
\frac{\partial \mbox{\boldmath $v_1$} }{\partial t }  = 
-\frac{1}{\rho_0} \mbox{\boldmath $\nabla$}  P_1 ,  
\end{equation}
\begin{equation}\label{3}
\frac{\partial}{\partial t} P_1 + F_p P_1 = \Gamma \frac{P_0}{\rho_0} (\frac{\partial}{\partial t}
\rho_1 + F_{\rho} \rho_1) ,  
\end{equation}
where, in general, the perturbed quantities are written as $\rho =
\rho_0 + \rho_1$, $\mbox{\boldmath $v$} = \mbox{\boldmath
$u$}+\mbox{\boldmath $v_1$}$, and $P = P_0 + P_1$.  The energy equation
is written following Hunter \& Whitaker (1989) with
$$
F_p = \frac{\Gamma (\Gamma -1)}{\rho_0 a^2} C_0 \left(
\left. \frac{\partial \ln C}{\partial \ln T}  \right|_{\rho} -
\left. \frac{\partial \ln H}{\partial \ln T} \right|_{\rho} \right), 
$$
and 
$$
F_\rho = \frac{\Gamma -1}{\rho_0 a^2} C_0 \left( 
\left. \frac{\partial \ln C }{\partial \ln T}\right|_\rho - 
\left. \frac{\partial \ln C}{\partial \ln \rho} \right|_T + 
\left. \frac{\partial \ln H}{\partial \ln \rho} \right|_T - 
\left. \frac{\partial \ln H}{\partial \ln T} \right|_\rho  
\right) , 
$$ 
where $C_0$ is the initial cooling rate, $a \equiv (\Gamma
P_0/\rho_0)^{1/2}$ is the sound speed, and $\Gamma$ is the adiabatic
index. 

In cylindrical geometry a random perturbation of $\rho _1$, ${\bf
v}_{1}$, and $P_1$ to an initial equilibrium state $\rho _0$, ${\bf
u}$, and $P_0$ can be considered to consist of Fourier components
of the form
\begin{equation}\label{4}
f_1(r,\phi ,z)=f_1(r)\exp [i(kz\pm n\phi -\omega t)] 
\end{equation}
where flow is along the $z$-axis, and $r$ is in the radial
direction with the flow bounded by $r=R$. In cylindrical geometry $k$
is the longitudinal wavenumber, $n$ is an integer azimuthal wavenumber,
for $n>0$ the wavefronts are at an angle to the flow direction, the
angle of the wavevector relative to the flow direction is $\theta =\tan
(n/kR)$, and $+n$ and $-n$ refer to wave propagation in the clockwise
and counterclockwise sense, respectively, when viewed outwards along
the flow direction. In equation (4) $n=$ 0, 1, 2, 3, 4, etc. correspond
to pinching, helical, elliptical, triangular, rectangular, etc. normal
mode distortions of the jet, respectively.  Propagation and growth or
damping of the Fourier components is described by a dispersion
relation
\begin{equation}\label{5}
\frac{\beta_{jt}}{\chi_{jt}} 
\frac{J'_n (\beta_{jt} R_{jt})}{J_n (\beta_{jt} R_{jt})} = 
\frac{\beta_{ex}}{\chi_{ex}}  
\frac{H'_n(\beta_{ex} R_{jt})}{H_n(\beta_{ex} R_{jt})} , 
\end{equation}
where the primes denote derivatives of the Bessel ($J$) and Hankel ($H$)
functions with respect to their arguments, $\beta$ and $\chi$ are
given by 
$$
\beta_{ex} = [-k^2 + \frac{\omega^2}{a_{ex}^2} Q_{ex} ] ^{1/2} , 
$$
$$
\beta_{jt} = [-k^2 + \frac{(\omega - k u)^2}{a_{jt}^2} Q_{jt} ] ^{1/2} ,
$$
and
$$
\chi_{ex} = \rho_{ex} \omega ^2 ,
$$
$$ 
\chi_{jt} = \rho_{jt} (\omega - k u)^2 .
$$
All heating and cooling information is contained in the dimensionless terms 
$Q_{jt}$ and $Q_{ex}$ where 
$$
Q_{ex} = \frac{1 + i F_p^{ex}/ \omega}{1 + i F_\rho^{ex}/ \omega} ,    
$$
and
$$
Q_{jt} = \frac{1 + i F_p^{jt}/ (\omega-ku)}{1 + i F_\rho^{jt}/ (\omega-ku)} .
$$
For normal mode $n$ the axial wavelength associated with a $360^{\circ
}$ helical twist of a wavefront around the jet beam is given by
$\lambda _z=n\lambda _n$ where $\lambda _n=2\pi /k$. The angular
frequency appearing in the linear analysis, $\omega$, represents a
$360^{\circ }/n$ helical twist and is properly related to an angular
precession frequency by $\omega = n\omega_p$.  For example, if an
elliptical jet distortion, normal mode $n=2$, rotates $360^{\circ }$
with the precession frequency $\omega_p$, at a fixed azimuthal angle
the frequency at which the jet surface oscillates is $2\omega_p$. In
general, each normal mode, $n$, consists of a single ``surface'' wave
and multiple ``body'' wave solutions to the dispersion relation.
Structural differences between the surface and body waves are discussed
in \S 3.4.

\vspace{-0.6cm}

\subsection{Heating and Cooling Rates}

Following previous 2D work (\cite{hs97}; \cite{sxh97}), we chose a jet
of radius $R_{jt} = 2.5 \times 10^{15}$ cm with number density $n_{jt}
= 600$ cm$^{-3}$, temperature $T_{jt} = 10^3 $ K and sound speed
$a_{jt} = 3.73 \times 10^5$ cm s$^{-1}$. Initially the jet is in
pressure equilibrium with an external (cocoon) gas with number density
$n_{ex} = 60$ cm$^{-3}$, temperature $T_{ex} = 10^4$ K and sound speed
$a_{ex} = 1.18 \times 10^6$ cm s$^{-1}$. With heating and cooling rates
of the form $H=\Lambda _Hn^{\alpha _H}T^{\beta _H}$, and $C=\Lambda
_Cn^{\alpha _C}T^{\beta _C}$, F$_p$ and F$_\rho $ are given by
$$
F_p=\frac{\Gamma (\Gamma -1)}{\rho _0a^2}{\bf C}\left( \beta _C-\beta
_H\right) \text{,} 
$$
and
$$
F_\rho =\frac{\Gamma -1}{\rho _0a^2}{\bf C}\left[ (\beta _C-\beta
_H)+(\alpha _H-\alpha _C)\right] \text{,} 
$$
where ${\bf C}=\Lambda _Cn_0^{\alpha _C}T_0^{\beta _C}$ ergs cm$^{-3}$ s$
^{-1}$. We will assume that the heating rate is independent of the
temperature, i.e., $\beta _H=0$, and is proportional to the density, i.e., 
$\alpha _H=1$, and the initial equilibrium requires that $H_0=C_0$.

In the linear analysis a piecewise power-law fit to the DM cooling function is represented by
$$
\Lambda_{DM} = \left\{ 
\begin{array} {r@{\quad:\quad}l}       
3.0 \times 10^{-28} T^{0.5} {\rm ~erg~cm}^{3}{\rm ~s}^{-1} & T < 10^4 {\rm ~K} \\ 1.5 \times 10^{-24} T^{0.5} {\rm ~erg~cm}^{3}{\rm ~s}^{-1} & T \ge 10^4 {\rm ~K} 
\end{array} 
\right. 
$$ 
For DM cooling, ${\bf C}=\Lambda _Cn_0^{\alpha _C}T_0^{\beta _C}=n_0^2 \Lambda_{DM}$ and for our choice of parameters
$$
\QATOPD\{ \} {F_p^{ex}}{F_p^{jt}}=\frac 5{9\rho _0a^2}\Lambda
_Cn_0^2T_0^{0.5}=\QATOPD\{ \} {2.16\times 10^{-9}{\rm ~s}^{-1}}{1.37\times
10^{-11}{\rm ~s}^{-1}}\text{,} 
$$
and
$$
\QATOPD\{ \} {F_\rho ^{ex}}{F_\rho ^{jt}}=-\frac 1{3\rho _0a^2}\Lambda
_Cn_0^2T_0^{0.5}=\QATOPD\{ \} {-1.29\times 10^{-9}{\rm ~s}^{-1}}{-0.82\times
10^{-11}{\rm ~s}^{-1}}\text{.} 
$$

\noindent
A power-law fit to MB cooling curve of the form
$$
\Lambda_{MB} = 1.5 \times 10^{-34} T^{2.53} {\rm ~erg~cm}^{3}{\rm ~s}^{-1}
$$
is used in the linear analysis to simulate the MB cooling curve. For MB cooling, ${\bf C}=\Lambda _Cn_0^{\alpha _C}T_0^{\beta _C}=n_0^2 \Lambda_{MB}$ and for our choice of parameters
$$
\QATOPD\{ \} {F_p^{ex}}{F_p^{jt}}\approx \frac{25}{9\rho _0a^2}\Lambda
_Cn_0^2T_0^{2.53}=\QATOPD\{ \} {1.46\times 10^{-10}{\rm ~s}^{-1}}{4.30\times
10^{-11}{\rm ~s}^{-1}}\text{,} 
$$
and
$$
\QATOPD\{ \} {F_\rho ^{ex}}{F_\rho ^{jt}}\approx \frac 1{\rho _0a^2}\Lambda
_Cn_0^2T_0^{2.53}=\QATOPD\{ \} {5.24\times 10^{-11}{\rm ~s}^{-1}}{1.54\times
10^{-11}{\rm ~s}^{-1}}\text{.} 
$$

The most important difference in jet stability properties between DM
cooling and MB cooling arises from the change in sign of $F_\rho $
(\cite{hs97}).  The positive value of $F_\rho $ for MB cooling is a
consequence of the steeper temperature dependence of the MB cooling
function. Positive values of $F_\rho $ occur when $(\beta _C-\beta
_H)+(\alpha _H-\alpha _C)>0$. For our choice of heating rate depending
linearly on the density and of cooling rate depending on the density
squared, $F_\rho >0$ when $\beta _C>1$. The DM power-law fit used in
the linear analysis serves to illustrate the effect of a shallow
dependence of radiative cooling on temperature in both jet and external
fluid. The MB power-law fit used in the linear theory serves to
illustrate the effect of a steep dependence of radiative cooling on
temperature in both jet and external fluid.  In the simulations and in
the theory the ``equilibrium'' heating rate $H_0 = C_0$ is determined
by the requirement that the jet and external medium be in thermal
equilibrium initially.  Because cooling rates are different in the jet
and external fluids as a result of temperature and density differences,
the initial heating rate required to establish and maintain thermal
equilibrium is different in the two fluids.

\vspace{-0.6cm}

\subsection{Dispersion Relation Solutions}

Numerical solutions to the dispersion relation for adiabatic (AD),
Dalgarno \& McCray (DM) cooling and MacDonald \& Bailey (MB) cooling
jets have been obtained for the pinch, helical, elliptical, triangular
and rectangular normal modes for $M_{ex} \equiv u/a_{ex} =$ 5 and 20
jets, jet speeds of 59~km~s$^{-1}$ and 236~km~s$^{-1}$, respectively.
The solutions for  $M_{ex} \equiv u/a_{ex} =$ 5 and 20 are
qualitatively similar, and in Figure 1 we show those appropriate to
$M_{ex} =$ 20 and a jet speed of 236~km~s$^{-1}$.  In the numerical
simulations, and possibly in protostellar systems, jets are perturbed
at their origin by a periodic motion at some frequency; thus, we solve
the dispersion relation for a complex wavenumber as a function of
angular frequency.  A negative value for the imaginary part of the
wavenumber indicates spatial growth at that angular frequency with an
e-folding length $\ell_e = \vert k_I^{-1} \vert$.

We identify the following features in the 3D solutions:

\noindent
(1) The  solutions for the pinch and helical normal modes in 3D are
very similar to the solutions obtained in 2D for the symmetric (pinch)
and asymmetric (sinusoidal) normal modes of the slab jet (\cite
{hs97}).  The higher order normal modes of the 3D jet which have no
analog on the 2D jet, e.g., the elliptical etc. normal modes, can be
thought of as harmonics of the helical mode and behave similarly to
helical mode solutions.

\noindent 
(2) In general, the linear growth rates, $\vert k_I \vert$, scale
inversely with the Mach number (not shown), and the linear growth
rates, especially for the body modes, are relatively smaller in 3D than
in 2D. Body wave mode growth rates can be larger, comparable to or
smaller than the growth rate of the surface wave mode for the pinch,
helical and elliptical normal modes, respectively. Body wave mode
growth rates are less than the surface wave modes for all higher order
normal modes (not shown).

\noindent 
(3) The effect of radiative cooling on the dispersion relation is
similar in 2D and 3D, i.e., DM cooling significantly increases the
linear growth rate at higher frequencies and MB cooling decreases the
linear growth rate at higher frequencies. The high frequency growth
rate plateau for DM cooling extends to frequencies only slightly higher
than those shown in the figure, at which point the growth rate rapidly
declines towards the adiabatic results at high frequency. Adiabatic and
MB cooling jets show distinct ``resonant'' (fastest growing) angular
frequencies, $\omega^*$, and wavelengths, $\lambda^*$, for almost all
surface and body wave solutions, and the DM cooling jet shows distinct
resonances for the body wave solutions.  As was found in 2D the
presence of DM cooling adds a pinch cooling mode (Ps2) that does not
exist for AD or MB cooling jets. In general, the wave modes are purely
real on the adiabatic jet or damped on the radiatively cooling jets
(damping rates not shown) in regions where they are not growing.

\vspace{-0.6cm}

\subsection{Fluid Displacements, Velocity \& Pressure Fluctuations}

Displacements, {\boldmath $\xi$}$(r,\phi ,z)$, of jet fluid from an
initial position $(r,\phi ,z)$ can be written in the form
\begin{equation}\label{6}
\text{\boldmath $\xi$}(r_0,\phi _s,z_s)={\bf A}(r_0)e^{i{\bf \Delta }(r_0)}\xi
_{r}(R)\exp [i(kz_s\pm n\phi _s-\omega t)]\text{ ,} 
\end{equation}
where $z_s$ and $\phi_s$ are the axial and azimuthal positions at the
jet surface, $r_0$ is the initial radial position, $\xi _{r}(R)$ is the
radial displacement at the jet surface, the ${\bf A}(r)e^{i{\bf \Delta
}(r)}$ are given by equations (A10) in Hardee, Clarke, \& Rosen (1997,
hereafter HCR), and the $k( \omega )$ are normal mode solutions to the
dispersion relation.  Fluid displacements are modified in amplitude and
rotated in azimuthal angle or shifted along the jet axis relative to
those at the jet surface by ${\bf A}(r)e^{i{\bf \Delta }(r)}$.  The
difference in structure between surface and body wave solutions on
adiabatic jets at their resonant frequencies as revealed by
displacement surfaces is shown explicitly by Figure 13 in HCR.  The
accompanying velocity perturbation, ${\bf v}_1(r,\phi ,z)=d${\boldmath
$\xi$}$/dt$, and pressure perturbation, $P_1(r,\phi ,z)$, can be
written in the form
\begin{equation}\label{7}
{\bf v}_1(r',\phi',z')=-i(\omega -ku){\bf A}(r_0)e^{i{\bf \Delta }(r_0)}\xi
_{r}(R)\exp [i(kz_s\pm n\phi _s-\omega t)]\text{ ,} 
\end{equation}
\begin{equation}\label{8}
P_{1}(r',\phi',z')=B(r_0)e^{i\Delta_p (r_0)}\xi _{r}(R)\exp [i(kz_s\pm
n\phi _s-\omega t)]\text{ ,} 
\end{equation}
where points on initially cylindrical surfaces are displaced radially
to $ r^{\prime }=r_0+\delta r$, axially to $z^{\prime }=z_s+\delta z$,
and azimuthally to $\phi ^{\prime }=\phi _s+\delta \phi $, and $\delta
r$, $\delta z$, $\delta \phi$ are the components of {\boldmath
$\xi$}$(r_0,\phi_s ,z_s)$.  In equation (8) $B(r_0)e^{i\Delta_p
(r_0)}=(\chi _{jt}/\beta _{jt})[J_n(\beta _{jt}r_0)/J_n^{\prime }(\beta
_{jt}R)]$ [e.g., \cite{hrhd98} eq.\ (14)]. The total velocity is given
by ${\bf v}={\bf u}+{\bf v}_1$ and the total pressure is given by
$P=P_0+P_1$.

If the azimuthal and axial phase shift is small then the radial fluid
displacement of a ``surface'' wave mode $n>0$ in the jet interior at
constant azimuthal angle is $ \xi _{r}(r) \approx \xi
_{r}(R)(r/R)^{n-1}$ (\cite{h83}).  The accompanying velocity and
pressure variations produced by higher order surface modes also show a
rapid decrease inwards.  On the other hand, at a constant azimuth the
``body'' wave modes have a reversal in fluid displacement at ``null
displacement'' surfaces in the jet interior and typically the maximum
pressure is near the null surface.

Jet distortion and accompanying pressure and velocity structure
associated with individual normal mode surface and body waves at
frequencies less than or equal to the resonant frequency have been
thoroughly investigated for adiabatic flows (\cite {hrhd98}; \cite
{h2000}).  Jet distortion and the accompanying pressure and velocity
structure across the jet and parallel to the jet axis are not strongly
modified by radiative cooling and the structure of the individual wave
modes on adiabatic and cooling jets as a function of frequency at and
below the resonant frequency of the appropriate mode is similar.  The
numerical simulations presented in the next several sections perturb
the jets at low frequencies and at high frequencies above the resonant
frequency.  At the higher frequencies the simulations reveal
interesting wave-wave interactions that have not been investigated
previously.  We will consider the high frequency structure of various
modes in comparison with structures observed in the numerical
simulations in \S 4.3.

\vspace{-0.6cm}

\section{Numerical Simulations}

\vspace{-0.2cm}

\subsection{Initial and Boundary Conditions}

All of the simulations were performed using the modified
three-dimensional hydrodynamic code CMHOG. For details on the
hydrodynamical algorithms see Stone, Xu, \& Hardee (1997). At
dimensionless simulation time $\tau_{sim} \equiv (a_{ex}/R_{jt})t = 0$,
a jet beam of Mach number $M_{ex}$ with uniform density $n_{jt}$ and
$z$-velocity $u_z  = M_{ex} a_{ex}$ is set up across a Cartesian
computational grid in the $z$ direction. Outflow boundary conditions
are used except where the jet enters the computational grid where
inflow boundary conditions are used.  The size of the computational
domain is varied in the axial direction depending on the Mach number of
the jet and the perturbation frequency in order to ensure that many
wavelengths of the most unstable mode are captured.  Simulations with
DM and MB type cooling were performed, along with simulations of
adiabatic jets (AD) that serve as a baseline for comparison.  Grid size
along with all the other key parameters associated with the simulations
are listed in Table 1.  Note that the duration of various simulations,
$\tau_{sim}$, is different.

In these simulations we study the dynamical evolution of unstable jets
that are initially in a delicate pressure equilibrium with a
low-density ambient medium and in which the gas is established in
thermal equilibrium with net cooling rate  $n^2\Lambda - nH = 0$.  To
maintain thermal equilibrium a different heating rate, $H$, is used in
jet and ambient fluid and the initial equilibrium could be maintained
for many dynamical times if the system was not perturbed.  As the
perturbed simulations evolve and the fluids mix the heating rate, $H$,
is varied according to whether a grid zone temperature is above or
below 10,000~K for DM cooling or according to the fraction of ambient
and jet material in a grid zone for MB cooling.  These different
approaches reflect the discontinuity in the DM cooling function at
10,000~K and the continuity of the MB cooling function, and are
identical to the methods used in the previous 2D simulations.  The per
particle cooling rate, $\Lambda$, is computed in each grid zone
according to the complete DM or MB cooling curves (see Figure 1 in
\cite {sxh97}).  Note that the net cooling rate is now non-zero but is
less than $n^2\Lambda$.  As a result the cooling length $\ell_{cool} >
v_{shock}t_{cool} \sim v_{jt}t_{cool}$ where $t_{cool} \equiv (\Gamma -
1)^{-1}(kT/n\Lambda)$, and shocked regions in the ambient medium can be
underesolved with cooling length a few times the grid spacing.
However, underesolution of these shocks does not influence the
stability or internal structure of the jet beam which is primarily what
we study here.

A periodic precession of the jet velocity is applied at the jet nozzle
to break the symmetry.  In the simulations, the pitch angle, $\theta$,
of the jet velocity relative to the axial direction varies from $\theta
=$ 0.0025 to 0.01 rad depending on the precession frequency.  As in
previous 2D simulations (\cite {sxh97}) we choose the initial precession
(perturbation) frequency to be $\omega R_{jt}/u =$ 0.1, 0.4 and 1.0 for
Mach 5 jets, and $\omega R_{jt}/u =$ 0.025 and 1.0 for Mach 20 jets.
With our assumed jet radius and velocities the precession period ranges
from about 840 to 21 years.

\vspace{-0.6cm}

\subsection{Surface Wave Modes}

\vspace{-0.2cm}

\subsubsection{Low Frequency Precession}

We performed simulations with angular frequency $\omega R_{jt}/u=0.1$
and $\omega R_{jt}/u=0.025$ on Mach 5 and 20 jets, respectively.  This
angular frequency is about a factor of three below the resonant
frequency on the AD jet, and for our assumed jet radius and flow speeds
corresponds to a precession period $\tau_p = 843.7$ yr.  Mach 20 jets
have speeds that are comparable to observed protostellar jets. As
comparable behavior is seen in the Mach 5 and 20 numerical simulations,
albeit with different length scale we show a volumetric rendering of
the density for Mach 20 AD and DM jets in Figure 2. At precession
frequencies much below the AD helical resonance, the linear theory
indicates that helical and higher order surface wave modes will be
dominant as the surface mode growth rates are much higher than the body
mode growth rates  on AD jets and body modes are damped on DM and MB
jets at this low precession frequency (see Figure 1).

The dominant oscillation is the result of helical twisting excited by
the precession at the inlet.  Note that the Mach 20 AD jet remains
collimated across most of the computational grid even though the linear
growth rates of AD and DM jets are nearly the same at this low
precession frequency.  Thus, non-linear processes speed the breakup of
radiatively cooling jets relative to adiabatic jets. A similar result
was found in 2D for identical parameters (see Figure 9 in \cite
{sxh97}). Some high density knot formation is apparent in the region
where the Mach 20 DM jet breaks up at a distance slightly less than
$400R_{jt}$, approximately 2.3 e-folding lengths. A 2D DM jet breaks up
at a slightly longer distance at this precession frequency even though
the 2D sinusoidal surface mode growth rate is larger  than the 3D
helical surface mode growth rate (about 50\% larger).  Upon break up
the 2D jet develops denser knots.  We note that the Mach 5 DM jet
develops similarly but only propagates about 1/4 the distance of the
Mach 20 jet -- about $100R_{jt}$ at break up.  This difference is about
the factor indicated by the direct scaling between growth length and
Mach number predicted from theoretically computed spatial growth
rates.  Very little knot formation occurs at the lower speed of the
Mach 5 jet in 2D or in 3D simulations.

In the region where Mach 20 and Mach 5 DM jets break up the structure
is very complicated and bow shocks are the dominant feature.  This
region also shows rapid mixing of jet and ambient material.  The
mixing, as measured by an entrainment volume and by an entrained mass,
for the Mach 5 DM simulation as a function of position along the jet
beam at time $\tau_{sim} = 100$ is shown in Figure 3. The entrainment
volume $V_{\epsilon}$ is defined as that volume of the fluid that has
$C > \epsilon$ where $0 < \epsilon < 1$ (e.g., \cite{lbbn96}), and $C$
is a ``color'' variable that traces the jet fluid.  A value
$V_{\epsilon} = 1$ indicates that all zones on the grid in a plane
transverse to the jet axis at location $z$ contain jet material with $C
> \epsilon$. The entrained mass $M_{\delta}$ is defined as the mass of
fluid with an axial velocity $v_z > \delta a_{ex}$ (e.g.,
\cite{lbbn96}; \cite{bw95}).  Mixing is considerably enhanced in 3D
relative to a similar 2D Mach 5 numerical simulation (see Figure 11 in
\cite{sxh97}).  The volume containing entrained material saturates near
$z = 50R_{jt}$ in 3D versus 300$R_{jt}$ in 2D. This is due to the
dramatic increase in area of the jet ambient medium interface when the
3D jet disrupts, thus accelerating the mixing process.  The entrained
mass shows an increase only for axial velocities of jet and entrained
material that are less than the external sound speed, i.e., $\delta <
1$, whereas in 2D the entrained mass showed an increase for $\delta
\leq 2$.  This is a result of the fast break-up and rapid slow down of
jet material accompanying the mixing process in 3D. Note that the
decline in entrained volume and mass at axial distances $z > 125
R_{jt}$ has occurred as flow from the inlet no longer reaches these
distances at time $\tau_{sim} = 100$.  Presumably the slowly moving
mixed material would progress across the computational grid at longer
times.

\vspace{-0.6cm}

\subsubsection{Moderate Frequency Precession}

In the Mach 5 simulations the angular precession frequency $\omega
R_{jt}/u=0.4$ is on the order of the resonant frequency of the first
few normal mode ``surface'' waves on the AD and MB jets, and is within
the unstable frequency range for the important accompanying ``body''
waves on AD, DM, and MB jets.  For our assumed jet radius and flow
speed this angular frequency corresponds to a precession period of
$\tau_p = 210.9$ yr.  No comparable simulations were performed for Mach
20 jets. The jets showed helical twisting at wavelengths,
$\lambda/R_{jt} \sim$ 14.3 (AD), 15.6 (DM), and 14.3 (MB), very close
to those predicted theoretically -- 14.1 (AD), 15.6 (DM) and 14.0 (MB)
-- where theoretical wavelengths are given by $\lambda/ R_{jt} = 2\pi
v_{gp}/\omega$ and $v_{gp} \equiv (\partial \omega / \partial k)
\vert_{real}$ is the group velocity at the precession frequency. Thus,
the linear analysis correctly predicts the helical wavelengths.

At this precession frequency the surface modes all have substantial
growth rates.  Ultimately, growth of these modes can cause the jet to
bifurcate or trifurcate. Splitting the jet into filaments provides a
large interface between the jet and ambient medium, and the observed
rapid mixing with the external medium.  Quantitative comparison between
linear growth rates shown in Table 2 indicates that the linear growth
rate for the helical surface mode is smallest for the adiabatic jet,
larger for the MB cooling jet, and largest for the DM cooling jet.  At
least qualitatively, the more rapid predicted growth of helical
instability on the cooling jets is verified by the simulations. The
rapid growth of higher order surface modes (see \S 4.2.4) in 3D
precludes a quantitative analysis of amplitude growth of the helical
mode. In any event, at a moderate precession frequency, radiative
cooling has accelerated the breakup of the jet beam, just as we found
at a low precession frequency.

Outside the jet beam the simulations show a pattern of spiral shocks
produced by the large amplitude helical twist.  In the external gas
this shock, when viewed edge on, appears as alternate bow-shaped arcs
on each side of the jet beam. This is illustrated in Figure 4 by a
volumetric rendering of the temperature and of the density from the
Mach 5 DM simulation.  High temperatures lie immediately behind the
shock front, and the volumetric rendering of the density shows that the
shocks originate at dense knots on the jet surface. In Figure 5, we
show a volumetric rendering of $\rho^2 T$ (density squared times
temperature) for this Mach 5 DM cooling jet. The $\rho^2 T$ image
emphasizes regions in the jet beam where the jet material is shocked
and raised to  higher temperatures and simulates emission from a single
line, e.g., [S~II].  At this frequency, the helical ridge on the
surface of the jet beam is caused by the  helical surface wave.  The
``hot spots'' are caused by projection combined with ``limb
brightening''. The high temperature postshock gas behind the spiral
shock forms a thin continuous radiating layer. If observed from  a
direction which is perpendicular to the jet beam, the radiative region
looks like discrete ``knots''.  At larger distances the jet starts to
entrain ambient material and dense knots are formed via cooling. These
knots have a velocity slightly less than the initial jet velocity and a
temperature which is much higher than the initial jet temperature.

\vspace{-0.6cm}

\subsubsection{High Frequency Precession}

The precession frequency $\omega R_{jt}/u=1.0$, about three times the helical
resonant frequency on the Mach 5 adiabatic jet, corresponds to a
precession period of $\tau_p = 84.4$ yr for our assumed jet radius and
flow speed. Figure 6 shows the jet density in a slice plane along the
jet axis for AD, DM, and MB Mach 5 jets at this frequency.  No
comparable simulations were performed for Mach 20 jets. The jet beams
do not show a large amplitude helical twist, but the internal patterns
reveal the presence of periodic knots inside the AD and DM jet beam.
This result is similar to numerical simulations in 2D slab geometry
(\cite{nh88}; \cite{hcc94}; \cite{sxh97}) which  have shown that the
jet beam cannot respond bodily to perturbations much higher than the
resonant frequency, i.e., the entire jet beam cannot form high
frequency sinusoidal or helical patterns with significant surface
displacement.

While these jets do not exhibit a large scale helical twist associated
with the surface wave at the precession frequency, the presence of a
resonant wavelength in the AD and MB jets is exhibited by a helical
twist seen in the AD and MB simulations as the jets break up.  In
particular, we measure a helical twist in AD and MB simulations with
wavelength $\lambda \approx$ $15R_{jt} \pm 2R_{jt}$. The linear theory
predicts a resonant wavelength for the helical surface wave of
$\lambda^* =$ $16.9R_{jt}$ and $13.2R_{jt}$ for AD and MB jets,
respectively, in good agreement with the simulations.  Thus, jet
breakup is associated with helical twisting at about the resonant
wavelength. These jets remain well collimated to distances in excess of
double that associated with the moderate precession frequency that was
near to the helical surface wave resonance. Thus, the collimation of
the jets is preserved fairly well due to the inability of jet beams to
respond bodily to the high precession frequency and very small
perturbation at the lower resonant frequency.

\vspace{-0.6cm}

\subsubsection{Higher Order Surface Modes}

Even though the precession that we use preferentially excites the
helical surface mode, the higher order surface modes have higher
maximum growth rates on AD, DM, and MB jets (see Figure 1) and can grow
to significant amplitudes.  Growth of the higher order modes is
observed to lead to elliptical, triangular, rectangular, etc.
distortions of the jet cross section in AD, DM, and MB Mach 5 jet
simulations.  This type of distortion is seen in all Mach 5 simulations
performed with precession frequencies $\omega R_{jt}/u =$ 0.4 \& 1.0.
The lower of these two precession frequencies is at about the
calculated resonant frequency for the helical through rectangular
surface modes on AD and MB Mach 5 jets, is slightly below the high
frequency growth rate plateau on the DM jet, and should excite these
surface modes at about their maximum growth rates. The higher
precession frequency lies in the range where body modes have their
maximum growth rates and where the surface modes have smaller growth
rates and (or for the DM jet with high frequency surface mode growth rate
plateau) have a lesser effect
on jet distortion.

The development of cross section distortion from the Mach 5 DM
simulation with precession frequency $\omega R_{jt}/u = 0.4$ is shown
in Figure 7. Qualitatively similar results are found for the comparable
Mach 5 AD and MB simulations.   As first observed by Hardee \& Clarke
(1995), we find that higher order modes with the fastest growth rates
appear first, and as one moves down the jet beam away from the nozzle,
the cross section area of the jet first becomes pentagonal, then
rectangular, triangular, and finally elliptical.  Note the jet beam
bifurcation as a result of the growth of the elliptical mode.
Quantitatively, the distance where we expect to see the higher order
modes to appear will be proportional to $\ell_e^* = \vert k_I^{*~-1}\vert$
where $k_I^*$ is the maximum spatial growth rate.  With our small
amplitude initial perturbation, growth to a significant amplitude
should require a distance of about $ 2\pi / k_I^*$, i.e., about six
e-folding lengths, and these distances are listed in Table 2.  We find
that rectangular, triangular, and elliptical distortions appear in the
same order along AD, DM and MB jets, and the locations of their
appearance are close to the values listed in Table 2.  Thus, the linear
analysis correctly predicts the relative growth rates of the surface
modes.

At a precession frequency of $\omega R_{jt}/u = 1.0$, the Mach 5 AD, DM
and MB jets remain well collimated to about twice the distance of their
$\omega R_{jt}/u = 0.4$ brethren.  Cross sections (not shown) reveal
lesser surface distortion but with internal structure consistent with
the development of body modes.  Thus, high frequency precession, where
the linear analysis predicts that surface mode growth is reduced and/or
perturbs only a shallow surface layer of the jet beam, is shown by the
numerical simulations to be much less destabilizing to the jet beam
than lower frequency precession as surface modes are suppressed.

\vspace{-0.6cm}

\subsection{The Body Modes \& Wave-Wave Interactions}

\vspace{-0.2cm}

\subsubsection{Low Frequency Precession}

At precession frequencies much below the AD helical resonance, the
linear theory indicates that helical and higher order surface wave
modes will be dominant, and the body wave modes are either purely real
on the AD jet or are damped on the DM and MB cooling jets.  While
growth may occur for all modes near resonance, the absence of initial
perturbations at higher frequencies suggests that growth of the surface
and/or body wave modes at resonance will be slow.  In fact we find
evidence for interaction between the helical surface wave and a weakly
damped helical body wave where both are excited by the precessional
perturbation.

An interaction between helical surface and body waves is shown in
Figure 8, which contains composite volumetric renderings of $\rho^2 T$
from the Mach 5 DM cooling jet perturbed at a frequency of $\omega
R_{jt}/u = 0.1$. The top panel shows the entire jet beam.  The middle
panel shows ``hot spots'' in the jet beam by excluding jet material
with temperatures at or below $1,000~K$.  The bottom panel shows an
enlargement of the knot region. The hot spots have temperatures ranging
from $1000~K - 10,000~K$ and appear very prominently at locations of $z
\approx$ 23 \& 37 $R_{jt}$. However, at the precessional frequency the
helical and higher order surface modes, have a wavelength of about $58
R_{jt}$ (from the linear analysis). Note the long wavelength
oscillation evident in the images.  On the other hand,  the  helical
body modes are only weakly damped at this frequency with damping length
$\sim 1000 R_{jt}$.  If the helical surface wave with wavelength
$\lambda_{surf} = 58.0 R_{jt}$, and the first helical body wave with
wavelength $\lambda_{body} = 18.47 R_{jt}$ are in phase at the origin,
subsequent in phase locations should occur at $23 R_{jt}$ and at $35.7
R_{jt}$.  These positions are very close to the observed positions of
the prominent knots at $z \approx$ 23 \& 37 $R_{jt}$.  The interaction
can also be seen in jet cross sections (not shown). Thus, the
interaction between body waves and surface waves can lead to knot
formation.  Knots or hot spots created in this fashion are stationary.

A second type of interaction also appears possible.  In the Mach 20 DM
jet perturbed with a precessional frequency of  $\omega R_{jt}/u =
0.025$ the jet ambient medium interface appears to break up at several
locations (see Figure 2 and compare the Mach 20 AD and DM jets in the
first half of the computational grid).  The break up of the jet-ambient
medium interface in this simulation appears to be the result of energy
deposition at the jet surface from a damped internal body wave.  With
the exception of this feature the Mach 20 DM jet shows very little
structure or evidence for dense knot formation inside the jet beam.

\vspace{-0.6cm}

\subsubsection{Moderate to High Frequency Precession}

The angular frequencies $\omega R_{jt}/u=$ 0.4 \& 1.0 are within the
unstable frequency range for the important ``body'' waves on AD, DM,
and MB Mach 5 jets.  In the DM and MB simulations at the lower
frequency, we observe dense filaments aligned with jet flow inside the
jet beam near to the surface, and see indications of dense knot
formation.  The location and length of these dense filaments is
coincident with constructive wave-wave interactions between the first
three helical body waves if all waves are launched from the origin in
phase.  For example,  density cuts in a slice plane along the jet axis
from the DM and MB simulations (not shown) reveal a dense filament of
length $\sim 4.1R_{jt}$ at $z \leq 10 R_{jt}$.  We note that full
wavelengths of the first, second and third helical body waves are
$\lambda^*/R_{jt}=$ 10.1, 7.5 and 6.0, and the first helical body wave
at $z = 10 R_{jt}$ has over-run the third helical body wave by a
distance of $\sim 4.1R_{jt}$ with less over-run for the second helical
body wave.  This suggests that compression over this length interior to
$z = 10 R_{jt}$ by the three body modes has led to radiative cooling
and the observed filament formation.

At the higher angular frequency of $\omega R_{jt}/u= 1.0$ density cuts
in a slice plane along the jet axis shown in Figure 6, and plots of
pressure, axial and transverse velocities along the jet axis, shown in
Figure 9 for the DM jet, reveal knot structure in the Mach 5 AD and DM
jets with periodic spacing of $34.0R_{jt}$ and $27.6R_{jt}$,
respectively.  Similar structure is not seen on the MB jet.  While the
structure suggests periodic pinching and knot formation, the periodic
spacing is much longer than the wavelength of any normal mode body or
surface wave at the precession frequency.  However, close examination
of ``resonant'' wavenumbers for first and second helical body modes
(Table 3) reveals a potential beat pattern with spacing of $39.2R_{jt}$
and $26.2R_{jt}$ for AD and DM jets, respectively, in excellent
agreement with the observed knot spacing.  The lack of a beat pattern
and knots in the MB jet might be  due to the low growth rate of the
second body wave relative to the first body wave on the MB jet.

How can periodic knots be produced by a beat pattern between helical
body modes?  Since only the beat pattern between the two helical body
modes can generate the observed wavelength, we interpret the simulation
results to mean that the wave-wave interaction between the helical body
waves has excited and funneled energy into a pinch mode at the long
beat wavelength.  To show how this could be so we have computed the
pressure and velocity fluctuations accompanying the first two helical
body modes, Hb1 and Hb2 using solutions computed from the dispersion
relation for the Mach 5 DM jet at the precession frequency $\omega R/u
= 1$, the fluctuations resulting from their beat pattern (Hb12), and the
fluctuations associated with the pinch surface cooling mode Ps2 at the
beat wavelength.  Fluctuations were computed using equations 6 -- 8 in
\S 3.4. The results for pressure fluctuations appropriate to those
observed in the Mach 5 DM simulation (Figure 9) are shown in Figure 10.
In Figure 10 we plot component structure along 1D cuts parallel to the
jet axis ($z$-axis) at different locations on the transverse $+y$-axis
where velocity components $v_x$ and $v_y$ represent azimuthal and
radial velocities, respectively.  Coarse mapping of values computed in
cylindrical ($r,\phi,z$) coordinates to the Cartesian ($x,y,z$)
coordinate location of the 1D cuts has led to the raggedness of the
lines in the figure.

The 1D cuts are shown at multiple radial locations between jet center
(dotted lines) and surface (dashed lines) for the first (Hb1) and
second (Hb2) helical body waves.  The highest pressures associated with
Hb1 at this frequency are near the jet surface.  The $v_y$ (radial)
velocity component shows opposing motions between material near the jet
surface and in the jet interior, and a half-wavelength axial phase
shift between outwards radial motion near the jet surface and at the
jet center.  The highest pressures associated with Hb2 are in the jet
interior with a half-wavelength axial phase shift in pressure extrema
between the jet surface and the jet center.  At this frequency the
radial velocity shows a large axial phase shift between jet surface and
jet center.  The axial velocity fluctuation is small for both helical
body modes near jet center. The combination of the two helical body
waves (Hb12) leads to a combined fluid displacement different from that
of the individual waves, and the resulting pressure and velocity
fluctuations along the jet axis (dotted lines) for the individual waves
cannot simply be added to give the pressure fluctuation (dash-dotted
line) and beat pattern in the transverse velocity components along the
jet axis.  Still the difference between the wavenumbers of the first
and second helical body modes (see Table 3) leads to the beat pattern
with wavelength of $ \sim 26R_{jt}$. The transverse velocity components
along the axis (dashed line \& dotted line) are $90^{\circ}$ out of
phase as expected for helical distortion.   Note the lack of
fluctuation in the axial velocity component as is observed in the
simulation.  Pressure and velocity fluctuation structure resulting from
this wave-wave interaction changes significantly off the jet axis.

The pressure and velocity fluctuations expected to accompany the pinch
cooling mode (Ps2) between the jet axis and surface (dash-dotted line)
at a wavelength comparable to the beat pattern between the two helical
body mode waves and with a comparable pressure fluctuation shows a
substantial axial velocity fluctuation and a lack of radial velocity
fluctuation along the jet axis (dotted line) as is observed in the
simulation.  Pressure and axial velocity fluctuation do not change as a
function of the radius but radial velocity fluctuation increases
slightly towards the jet surface (dashed line). Clearly wave-wave
interaction between the helical body modes couples via the pressure
fluctuation near jet center to the pinch cooling mode with pressure
fluctuation occupying a large fraction of the jet interior and with
some non-linear interaction between the velocity components that
results in both small axial and transverse velocity fluctuation near
jet center.  The resulting nearly axisymmetric periodic pressure and
density fluctuation produces the observed knots.

\vspace{-0.6cm}

\subsubsection{Very High Frequency Precession}

In the Mach 20 DM simulation we study jet response to an angular
precession frequency of $\omega R_{jt}/u=1.0$ that is over ten times
the AD jet helical resonant frequency. For our assumed jet radius and
flow speed this angular frequency corresponds to a precession period of
$\tau_p = 21.1$ yr.  At this frequency, the linear theory indicates a
high frequency plateau in the surface wave mode growth rate that is
above the body wave mode growth rates (see Figure 1).  The helical
nature of the initial precession guarantees that modes higher than
helical are unlikely to develop  to significant amplitude, although
growth rates are comparable.  We note that the wavelengths predicted
to accompany this precession frequency are shorter for the helical
surface mode than for helical body modes.  This is a reversal of the
usual relationship.

This simulation provides another interesting example of knot formation via
wave-wave interaction.  Knot formation appears in the $\rho^2 T$
volumetric rendering image in the upper panel in Figure 11 at
$\tau_{sim} = 5$.  The volumetric rendering reveals a clear helical
pattern within the jet beam with wavelength $6.5R_{jt}$. From the
linear analysis, we calculate wavelengths of $\lambda_h = $
4.8$R_{jt}$, 6.65$R_{jt}$, 6.31$R_{jt}$, and 6.05$R_{jt}$ for the
helical surface and first three helical body modes, respectively, at
the precession frequency.   Thus, we tentatively identify the dominant
pattern with the first helical body wave and the ``hot spots'' (at $z
\sim$ 20 \& 35 $R_{jt}$ in Figure 11) in the helical pattern as caused
by wave-wave interaction between the helical surface wave and the first
helical body wave as illustrated qualitatively by the lower panel in
Figure 11.  Additional information concerning wave-wave interactions is
provided by profiles of pressure, axial velocity, and transverse
velocity components along the $z$-axis shown in Figure 12.  These
profiles reveal that the transverse velocity components undergo out of
phase sinusoidal oscillation (indicating helical twisting at jet
center) which dies away and then returns at $z \sim 55R_{jt}$.   The
difference between the wavenumbers of the first and second helical body
modes (see Table 3) can lead to a beat pattern with wavelength of
$ \sim 120R_{jt}$ similar to that revealed in
Figure 12. While the pressure and axial velocity remain constant out to
$z=25R_{jt}$, Figure 12 shows the development of quasi-periodic
pressure and axial velocity oscillation with an observed wavelength
$\sim 6.7 R_{jt}$ at larger distance that can only be produced by the
second pinch body mode.

We show that the hot spots can be fit by interaction between the
helical surface (Hs) and first helical body (Hb1) waves, and that the
structure at jet center can be fit by a beat pattern between first
(Hb1) and second (Hb2) helical body modes combined with the presence of
the second pinch body mode (Pb2) in Figure 13.  In Figure 13
fluctuations were computed using equations 6 -- 8 in \S 3.4. Component
structure is shown along 1D cuts parallel to the jet axis ($z$-axis) at
different locations on the transverse $+y$-axis where velocity
components $v_x$ and $v_y$ represent azimuthal and radial velocities,
respectively, like Figure 10.  The pressure and velocity fluctuations
that would accompany the helical surface (Hs), first body (Hb1) and
second body (Hb2) waves using solutions computed from the dispersion
relation for the Mach 20 DM jet at the precession frequency $\omega R/u
= 1$ are shown.  This frequency is above the resonant frequency for both body
modes and lies on the high growth rate plateau for the surface mode
wave.  The axial phase shift in pressure and radial velocity
fluctuations between jet center and surface for Hb1 is much smaller
than that shown in Figure 10 for the Mach 5 solution.  The difference
is entirely the result of the difference in the frequency relative to
the resonant frequency for this body mode.  The axial phase shift in
pressure and velocity fluctuation between jet center and surface for
Hb2 is similar to that shown in Figure 10 for the Mach 5 solution.
Note that significant pressure and velocity fluctuations
associated with the surface wave occur only near to the jet surface (dashed
line).  Thus, the surface wave does not influence the jet interior at
this high frequency. This result supports the conclusion that at high
frequencies the helical surface wave mode can effect only a shallow
layer near to the jet surface (\cite{hn88}; \cite{hs97}), whereas the
body waves can still effect a larger volume of the jet beam.

Figure 13 also shows the result of wave-wave interaction between the
surface and first two helical body modes (Hsb12). The  beat wavelength
(best illustrated by the transverse velocity fluctuations along the jet
axis) is comparable to that seen in the simulation.  Additionally, the
figure shows the pressure fluctuation at the jet surface (dashed line)
associated with the wave-wave interaction between the surface and first
two helical body modes.  Here we note an additional beat
wavelength that arises between the surface and body modes near the jet
surface, where the pressure fluctuations achieve maxima at axial
distances $z/R \approx$ 20, 42 and 65.  This confirms the tentative
identification of the hot spots in the simulation with interaction
between surface and body helical waves.  Note that the beat wavelength
shows a slow pressure variation along the jet axis (dotted line).  A
rise in the pressure along the axis occurs as the transverse
($v_x$,$v_y$) velocity fluctuations decrease.  This behavior is
identical to that seen in the body mode wave-wave interaction (Hb12)
shown in Figure 10. There is a small accompanying axial velocity
variation on the axis.  Just off the axis a short wavelength pressure
and axial velocity fluctuation (solid line) is evident. This short
wavelength fluctuation can couple to the pressure and velocity
fluctuation expected to accompany the second pinch body mode (Pb2) to
produce the short wavelength quasi-periodic pressure and axial velocity
oscillation seen in the simulation with an observed wavelength $\sim
6.7 R_{jt}$.  Thus the simulation results again show that a wave-wave
interaction between helical body modes can excite a pinch mode.

\vspace{-0.6cm}

\section{Summary}
    
\vspace{-0.2cm}

\subsection{Comparison with theory}

We have found that the analytical linear analysis provides both good
qualitative and quantitative agreement with our 3D simulations. The
linear analysis successfully indicates: (1) the dominate modes in
various perturbation frequency ranges, (2) the increase in instability
due to cooling, and (3) the frequency range in which mode-mode coupling
is most likely to occur. Quantitative measurements of parameters such
as the wavelengths of individual modes and the beat patterns formed by
wave-wave interaction are correctly predicted and the order of
occurrence of higher order surface modes are also matched
quantitatively with the predictions from the linear analysis.  In
addition, velocity and pressure fluctuations computed from the theory
appear to be in good quantitative agreement with fluctuations observed
in the numerical simulations.  The coupling between normal wave modes
that we have found here has also been found in numerical simulations
using a spectral type code by Keppens \& T\'oth (1999) who also have
identified analytically the quasi-linear coupling mechanism that
connects a driven helical mode to several other low order normal
modes.

\vspace{-0.6cm}

\subsection{Disruption of the jet beam}

3D simulations confirm results and conclusions reached in 2D
simulations (\cite{sxh97}). Cooling (especially DM type cooling)
increases the linear growth rate of unstable modes, especially the
helical surface and body modes, and helps to break up the jet beam
particularly  at perturbation frequencies at or below resonance
relative to the adiabatic jet. What is unique in 3D is that higher
order K-H modes play important roles in the jet evolution,
filamentation and disruption.  Animations of our 3D simulations and
Figures 2, 5, 6 and 7 show that disruption of the jet beam can be
categorized by two different fundamental processes: (1) the amplitude
of distortion caused by helical and other low order surface wave modes
grows large and the jet beam breaks into filaments in the direction of
jet flow. The increase in the contact area between the jet beam and the
ambient gas leads to enhanced mixing and leading edge shock and knot
formation, and (2) the compression wave front established inside the
jet by the initial perturbation is not reflected back into the jet beam
from the jet-ambient interface but deposits energy at the jet surface
and opens a gap in the jet surface, e.g., by a damped helical body
mode. The gap creates a shock in the ambient gas, and the jet is
disrupted.

It is clear that development of jet distortion, knot formation and jet
disruption depend on the radiative cooling rate and the perturbation
frequency.  When the development of jet distortion is slowed, by
whatever process, disruption is delayed.  In general, radiative cooling
speeds those processes leading to jet disruption relative to an
adiabatic jet.  DM cooling with a relatively shallow dependence of
radiative cooling on temperature in both jet and external fluid is
somewhat more destabilizing than MB cooling with a steep dependence of
radiative cooling on temperature in both jet and external fluid.  Jets
perturbed at high frequency even though helical surface and body modes
have large growth rates are relatively stable to disruption. It is the
value of the perturbation frequency relative to the resonant frequency
and not the absolute value that is important.  Typically, our jets
remained collimated to twice the distance when perturbed at a high
frequency when compared to jets perturbed at the resonant or lower
perturbation frequency.

\vspace{-0.6cm}

\subsection{Knots \& Shock Spurs}

We note that knots in the present 3D simulations were less dense than
in comparable 2D simulations.  Mode-mode interactions, also present in
2D simulations, are more complicated in 3D. The interactions, such as
helical surface and helical body mode interaction and coupling to pinch
modes, lead to new mechanisms for the formation of staggered emission
knots along the jet beam which can lead to staggered bow shocks or lead
to relatively stationary knots in the jet beam.

\noindent
(1) Shocks outside the jet beam created by large amplitude surface
displacement provide one mechanism for knot formation and for the
formation of shock spurs.   In this case, the jet-ambient interface
creates a spiral shaped shock around the jet beam in the ambient gas.
If viewed edge-on, this shock presents a staggered pattern of shock
spurs down the jet beam. Interior to the jet surface a strong density
enhancement near the surface of the jet beam provided by large
amplitude helical surface displacement accompanies the spiral shaped
shock.  Viewed edge-on the density enhancement looks like localized
knots at the base of the shock spurs.  These knots and shock spurs
move with the speed of the helical surface wave. 

\noindent 
(2) Wave-wave interactions provide a second mechanism for
knot formation.   In general, our results indicate some suppression of
knot formation at low precession frequencies resulting from the absence
of multiple body waves and the associated wave-wave interactions.  In
wave-wave interactions, the knots are formed internal to the jet beam
and in one simulation the knots are seen to be stationary in temporal
animations.  Thus, no knot driven shocks are created in the ambient
gas.   Note that the material moving through the stationary knots is
moving at nearly the jet speed and large radial velocities could still
be observed from the spectral lines.  However, in another simulation
the knots formed as a result of wave-wave interaction near to the inlet
are initially nearly stationary, but at larger distance the knots move
and develop shock spurs. At high frequencies surface distortion leading
to shock spurs does not occur.

\noindent
(3) As a jet breaks up mixing with the external environment leads to
shocks and knot formation. Knots formed in this fashion are primarily
associated with the bow-shock region at the head of a protostellar
jet.  In our simulations this region moves outwards into an already
outwards moving region where jet material previously existed.  As a
result the resulting bow shock is weaker than might be the case
otherwise.

\vspace{-0.6cm}

\section{Discussion}

Internal knots are one of the most prominent features of YSO jets.
Knots have been seen to form at the source and move outwards with time,
e.g., HH~80/81 (\cite{mrr95}) and the Serpens Radio Jets
(\cite{curiel95}). Since they bear many of the same spectral signatures
as individual HH objects, they should be related to radiative shocks.
In our study, all the initial perturbations are chosen to be fairly
small so as not to excite nonlinear dynamic effects.  As a result,
while we see complex internal structures we do not see strong shocks
internal to the jet beam as long as the jet is not disrupted. B\"uhrke,
Mundt, \& Ray (1988) and Bodo et al.\ (1994) have proposed that the K-H
pinch instability may cause knots to form in the jet beam.  Our
simulations do provide evidence for knots produced by pinching at
wavelengths induced by helical wave-wave interaction. In our
simulations the knots formed at the beat wavelength typically have
negligible proper motion. There are pulsating jet models (see
\cite{sn93b} and references therein) in which strong internal shocks
are created in the jet beam. Knots created by the pulsating jet will
have proper motions that are comparable to the jet speed.

Evidence for small and/or accelerating knot proper motion is provided
by detailed emission line observations of HH~83 by Reipurth (1989).
In  this  highly collimated jet there are about ten emission knots, and
the jet terminates at a strong  bow shock.  The distance to the source
is 460 pc (the source is in  the Orion Cloud), and the inclination
angle of the jet is $\sim 45^\circ$.  Thus, the physical extent of the
jet is about 0.1 pc.  From  the  observed width of the brightest
knots, the physical length-to-width ratio is at least 17.  Therefore,
from the source to the terminating bow shock the linear scale is at
least $30 R_{jt}$. Detailed spectral studies reveal  that there is a
systematic increase of the velocity from  knot A to knots I and J from
$-80$~km~s$^{-1}$ to $-180$~km~s$^{-1}$, and the bow shock has
H$_{\alpha}$ line emission only, which indicates a shock with velocity
of 80~km~s$^{-1}$ or greater.

These observational results may be compared to our numerical simulation
of a Mach 5 DM cooling jet (jet speed of about 60~km~s$^{-1}$)
perturbed with a precession frequency $\omega R_{jt}/u = 0.1$ (a period
of about 840 years) shown in Figure 8.  This volumetric rendering of
$\rho^2 T$ at an angle of inclination of $45^{\circ}$ is at simulation
time $\tau_{sym}= 100$ (a time of about 6700 years).  In the simulation
the linear scale of the knot forming region is $< 100 R_{jt}$ ($<
0.1$~pc). In the simulation the ``hot spots'' occur at the intersections
of helical surface and body waves.  The intersections interact with
the external medium and are shocked as the jet moves supersonically
through the ambient gas, and shock spurs develop in Figure 8 at knots
farther from the origin. Hot spots closer to the nozzle are relatively
``younger''and are just formed.  When a hot spot is newly formed as a
result of wave-wave interaction it should be nearly stationary.
However, the hot spot material moves with the jet flow and generates a
strong shock at the jet-ambient interface (the shock spurs in Figure
8) the material behind the shock spur should have a velocity of about
1/4 of the shock speed (also about 1/4 the jet speed).  Observed
emission should come from the higher temperature shocked material at
the jet ambient interface. As time passes, the knot is accelerated by
the jet and the gas in the knot mixes with the jet gas.  Thus, an
observation should show that a knot which is closer to the nozzle has
lower velocity, and knots farther from the nozzle, i.e., accelerated by
the jet, should move faster. Note also that some ``hot  spots'' are on
the far side the jet beam, which, as the helical wave develops, move
away from us along the line of sight with respect to the jet beam while
other hot spots are on the near side of the jet beam and move toward us
along the line of sight relative to the jet beam.  This mechanism
provides additional cyclic variation in the radial velocity of knots.

The initial transverse velocity perturbations used in our simulation
are quite small and cannot account for the large velocity variation
among the knots observed in HH~83.  However, the effect would be larger
if the initial perturbation had a considerable amplitude. The
termination of the observed jet after knot J would result from
disruption of the jet beam. Further downstream, the interaction of the
jet material with ambient gas would form the classic bow shock --
structure associated with the working surface observed in many HH
objects.  Because the morphology of the source and of the simulation
are very similar we believe that we have found a mechanism that could
result in variation or systematic increase in radial velocity of
knots.

Another example of similarity between observational morphology and our
simulations is provided by the structure of the jet in HH~111.  In
particular, the [S~II] images of the jet reveal a spiral emission
feature wrapped around the jet beam with embedded knots, and the
H$_{\alpha}$ images present us with remarkable staggered bow shocks
farther out along the jet beam (\cite{reipurth97}).  We can immediately
identify the spiral emission feature and bright knots with features
like those seen in the Mach 5 DM cooling jet perturbed with a
precession frequency of  $\omega R_{jt}/u = 0.4$ (a precession period
of about 210 years) or with the Mach 20 DM cooling jet (jet speed of
236~km s$^{-1}$) perturbed with a precession frequency of $\omega
R_{jt}/u = 1$ (a precession period of about 21 years).  Volumetric
rendering of $\rho^2T$ of the Mach 5 and Mach 20 simulations are shown
in Figures 5 \& 11, respectively.  In particular,  helical emission
bands and the knots seen in the Mach 20 simulation and in the [S~II]
map from Reipurth et al.\ (1997) have wavelength and spacing relative
to the jet radius that is similar in both the simulation and the
observed jet.  That this type of structure appears in both simulations
at very different precession periods and jet velocities suggests that
no fine tuning is required to generate such structure. Because of the
limited domain size in the Mach 20 simulation, we do not know whether
the observed staggered bow shock can be formed at larger distance.
However, the Mach 5 simulation that develops faster spatially suggests
that this would be so.

\vspace {1.0cm}

PEH acknowledges support from the National Science Foundation
through grant AST-9802955 to the University of Alabama.  JMS acknowledges
support from DOE grant DFG0398DP00215 to the University of Maryland.

\newpage

\begin{table} [ht]
\centering 
\caption{Physical and numerical parameters used in 3D simulations.}
\vspace{0.1cm} 
\begin{tabular}{c c c c c c c c c} \hline \hline
Mach \#  & Type$^{~1}$  & Grid  & z/$R_{jt}$ &  (xy)/$R_{jt}$ &
$\omega R_{jt}/u$ & $\theta (rad)$ & XYzones/$R_{jt}$ &  $\tau_{sim}$ \\ \hline 
20 & A,D & $200 \times 80 \times 80$ & 500 &$\pm 4$ & 0.025 & 0.0025
& 12 & 25 \\  
20 & D & $100 \times 80 \times  80$ & 60 &$\pm
4$ & 1.0 & 0.01 & 10 & 5 \\   
5 & D  & $200 \times 94 \times
94$ & 400 &$\pm 8$ & 0.1 & 0.0075 & 10 & 100 \\  
5 & A,D,M &
$200 \times 80 \times 80$ & 60 &$\pm 4$ & 0.4 & 0.03 & 10 & 15 \\
5 & A,D,M & $200  \times 120 \times 120$ &  300 &$\pm 10$ &
1.0 & 0.075 & 6 & 80 \\ \hline
\end{tabular}
\label{table1}
\end{table}

\vspace{-0.20cm}

$^1$ Cooling type -- A: Adiabatic, D: DM cooling, and M: MB cooling

\begin{table} [h!]
\vspace{-0.2cm}
\centering
\caption{Growth rates of surface modes for Mach 5 jets precessed at $\omega R_{jt}/u = 0.4$}
\vspace{0.1cm} 
\begin{tabular}{ l c c c c c c} \hline \hline
 Mode   &  $k_IR_{jt}$ (AD) & $k_IR_{jt}$ (DM) & $k_IR_{jt}$ (MB) & $2\pi /k_I$ (AD) & $2\pi /k_I$ (DM) & $2\pi /k_I$ (MB) \\ \hline  
 Helical$^{~1}$     &  $-0.062$ &  $-0.087$ & $-0.068$ & 101$R_{jt}$ & 72$R_{jt}$  & 92$R_{jt}$ \\
 Elliptical$^{~2}$  &  $-0.135$ &  $-0.273$ & $-0.131$ & 46$R_{jt}$  & 23$R_{jt}$  & 48$R_{jt}$ \\ 
 Triangular$^{~2}$  &  $-0.229$ &  $-0.274$ & $-0.210$ & 27$R_{jt}$  & 23$R_{jt}$  & 30$R_{jt}$ \\ 
 Rectangular$^{~2}$ &  $-0.336$ &  $-0.327$ & $-0.303$ & 18$R_{jt}$  & 19$R_{jt}$  &  21$R_{jt}$ \\ \hline 
\end{tabular}
\label{table2} 
\end{table}

\vspace{-0.20cm}

$^1$ Growth rate at precession frequency $\omega R_{jt}/u = 0.4$

\vspace{-0.20cm}

$^2$ Maximum growth rate 


\begin{table} [h!]
\vspace{-0.2cm}
\centering
\caption{Helical S, B1 \& B2 modes on jets precessed at $\omega R_{jt}/u = 1.0$}
\vspace{0.1cm}
\begin{tabular}{c c c c c c c c} \hline \hline
 Type  &  $k_RR_{jt}$ (S) & $k_RR_{jt}$ (B1) & $k_RR_{jt}$ (B2) & $\Delta k_R R_{jt}^{~1}$ & $k_IR_{jt}$ (S) & $k_IR_{jt}$ (B1) &  $k_IR_{jt}$ (B2) \\   \hline 
    & & & ~~~Mach 5 & Jets~~~~~ & & & \\

 AD & --- & 1.30 & 1.46 & 0.16 & --- & $-0.072$ & $-0.054$  \\
 DM & --- & 1.18 & 1.42 & 0.24 & --- & $-0.043$ & $-0.025$  \\ 
 MB & --- & 1.30 & 1.44 & 0.14 & --- & $-0.080$ & $-0.032$  \\

  & & & ~~Mach 20 & Jet~~~~~~ & & & \\

 DM & 1.3059 & 1.0339 & 1.0868 &  0.0529 & $-0.0572$ &  $-0.0012$ & $-0.0035$  \\ \hline
\end{tabular}
\label{table3}
\end{table}

\vspace{-0.2cm}

$^1$ [$\Delta k_RR_{jt}= k_RR_{jt}$ (B2) -- $k_RR_{jt}$ (B1)]

\newpage

\section{Figures}

\baselineskip 11 pt

\begin{figure}[h!]
\vspace{15cm}
\includegraphics{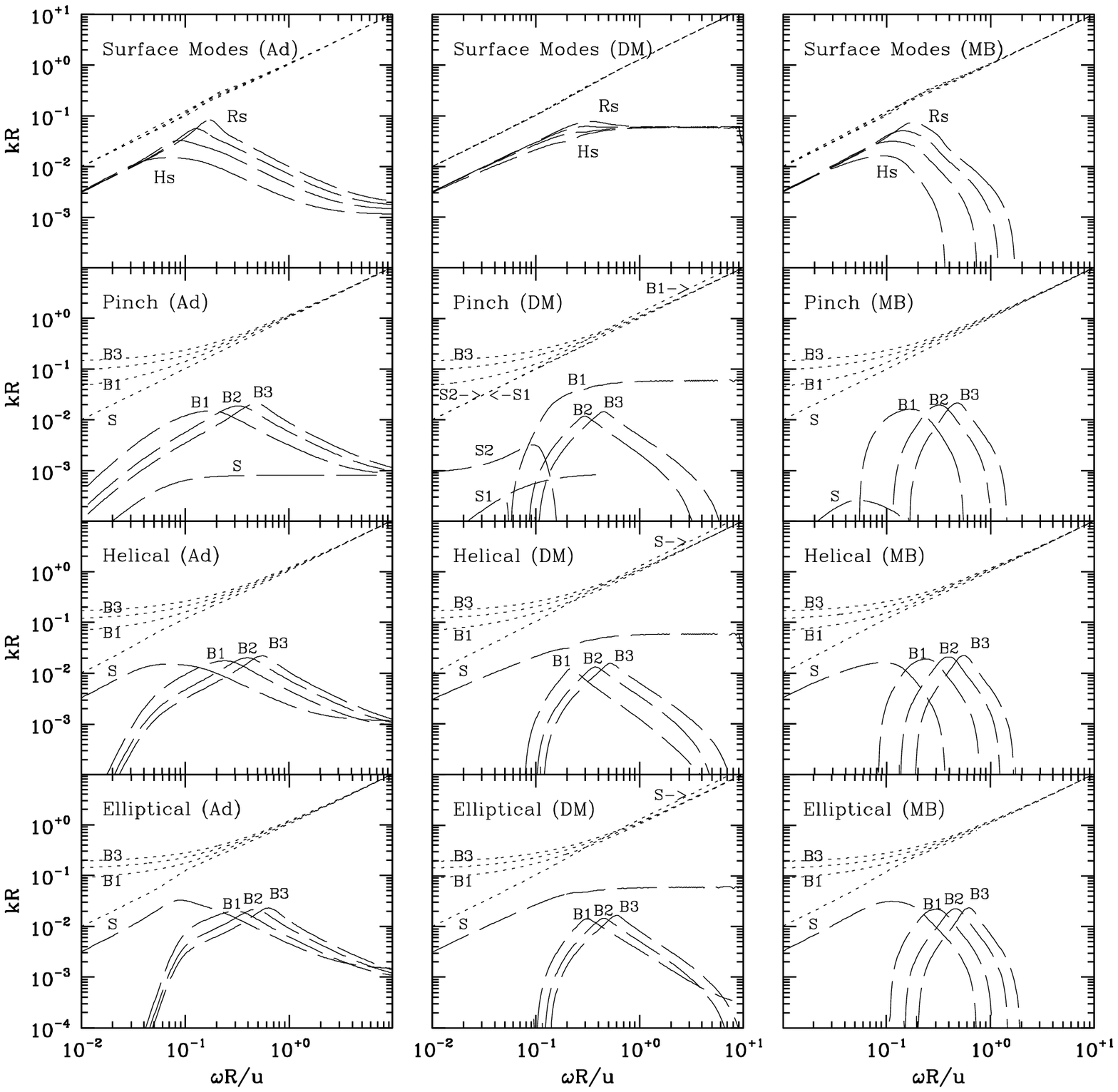}
\end{figure}

\noindent Fig. 1---Surface wave solutions to the dispersion relation for the
helical (Hs), elliptical, triangular and rectangular (Rs) normal modes
on the Mach 20 Ad(iabatic), DM cooling and MB cooling jets (top
panels). Surface (S) wave solutions and first three body wave (B1, B2,
B3) solutions to the dispersion relation for the pinch, helical and
elliptical normal modes on the Mach 20 Ad(iabatic), DM cooling and MB
cooling jets (lower panels). The dotted lines give the real part of the
wavenumber, $k_R$,  and the dashed lines give the absolute value of the
imaginary part of the wavenumber, $\vert k_I \vert$, as a function of
the angular frequency, $\omega$. Quantities are scaled by the jet
radius, $R_{jt}$, and the jet velocity, u.

\newpage

\begin{figure}[h!]
\vspace{4cm}
\includegraphics{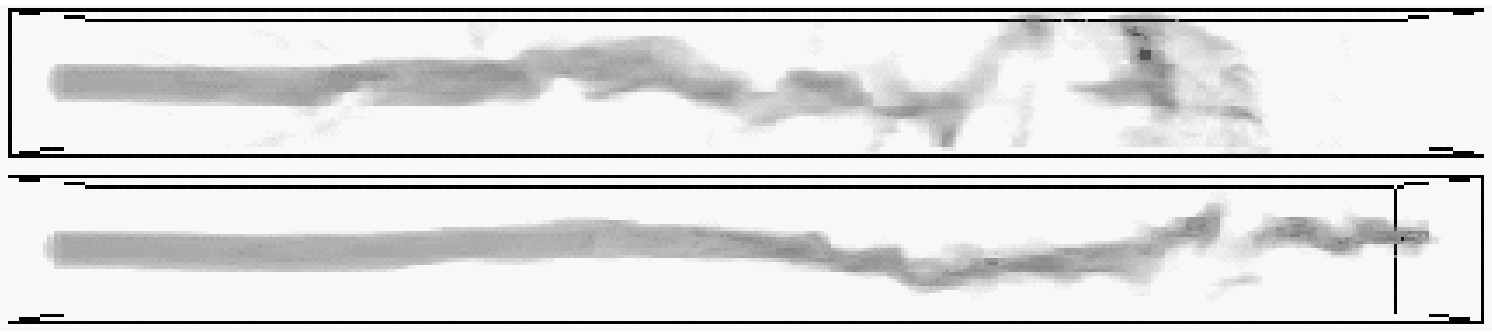}
\end{figure}

\noindent Fig. 2---Volumetric rendering of density for the Mach 20 DM jet
(top) and Adiabatic jet (bottom) with $\omega R_{jt}/u=0.025$.   The
volume shown is $500R_{jt} \times 8R_{jt} \times 8R_{jt}$. The images
are underscaled horizontally by a factor six. Darker shading indicates
higher density.

\begin{figure}[h!]
\vspace{9cm}
\includegraphics{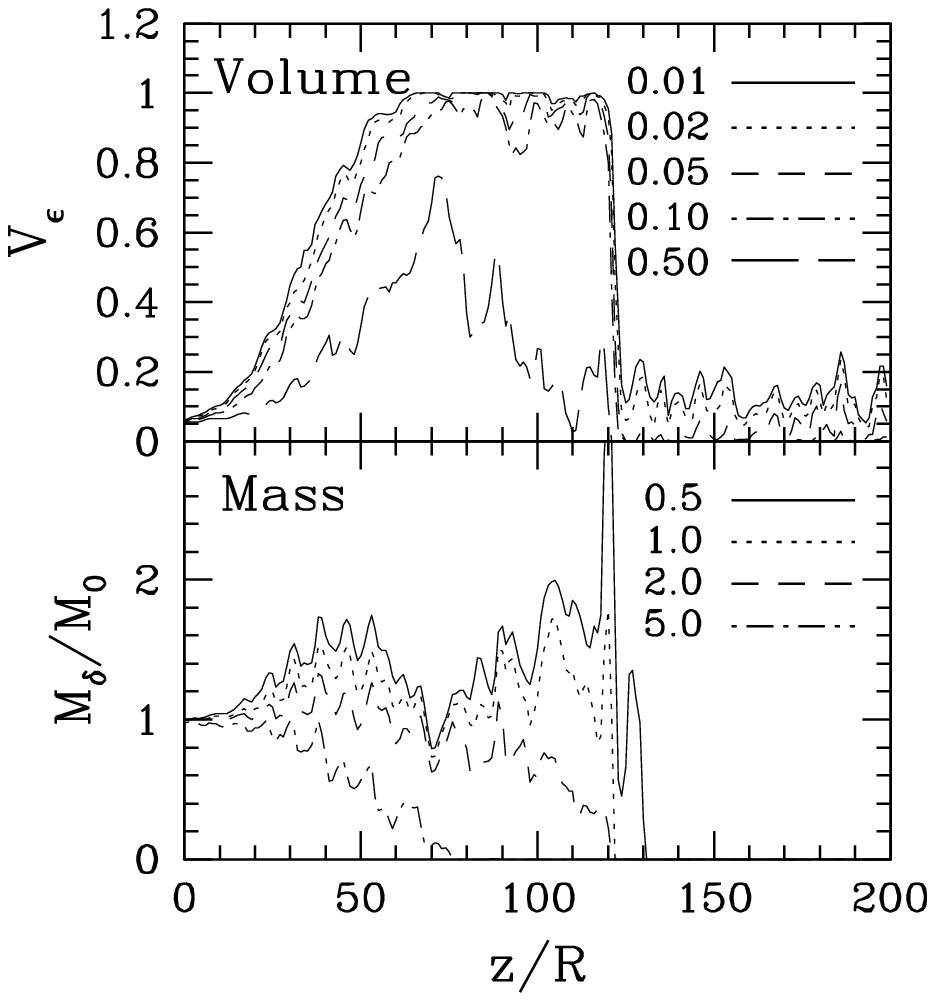}
\end{figure}

\noindent Fig. 3---Entrainment volume $V_{\epsilon}$ (top) and entrained mass
$M_{\delta}$ (bottom) vs. axial position for a Mach 5 DM jet with
$\omega R_{jt}/u=0.1$. Lines are at values of $\epsilon =$ 0.01, 0.02,
0.05, 0.1 \& 0.5. and $\delta =$ 0.5, 1, 2 \& 5.

\newpage

\begin{figure}[h!]
\vspace{6cm}
\includegraphics{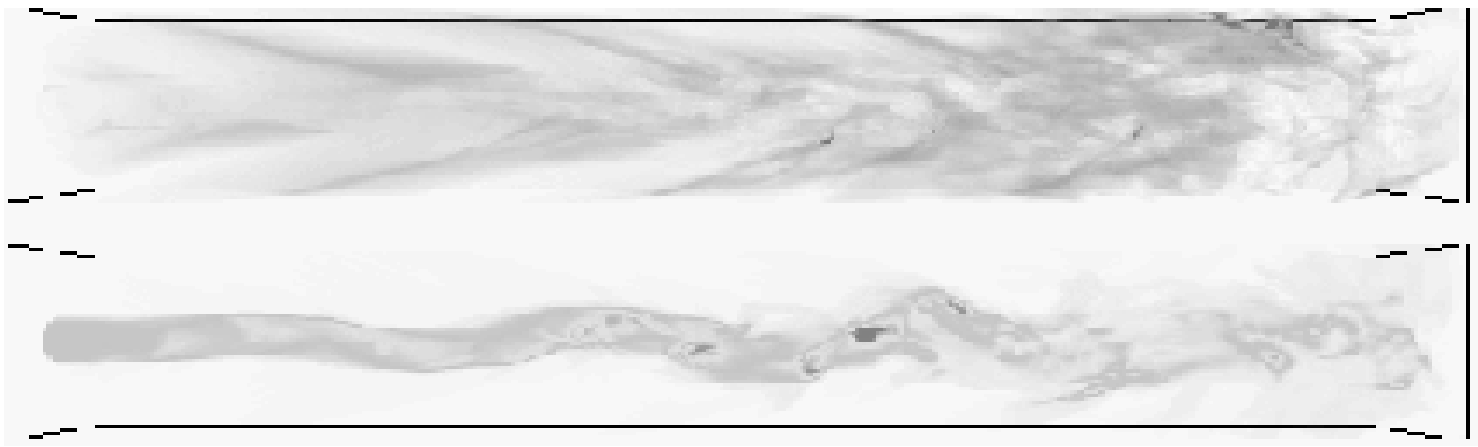}
\end{figure}

\noindent Fig. 4---Volumetric rendering of temperature (top) and density
(bottom) for a Mach 5 DM jet at $\omega R_{jt}/u=0.4$.  The maximum
temperature and density are $\approx 40,000$~K and $\approx
1,400$~cm$^{-3}$, respectively, and the volume shown is $60R_{jt}
\times 8R_{jt} \times 8R_{jt}$. Darker shading indicates higher
temperature.  In general higher densities are darker but in the first
half of the jet lighter areas are also above the initial jet density.

\begin{figure}[h!]
\vspace{4cm}
\includegraphics{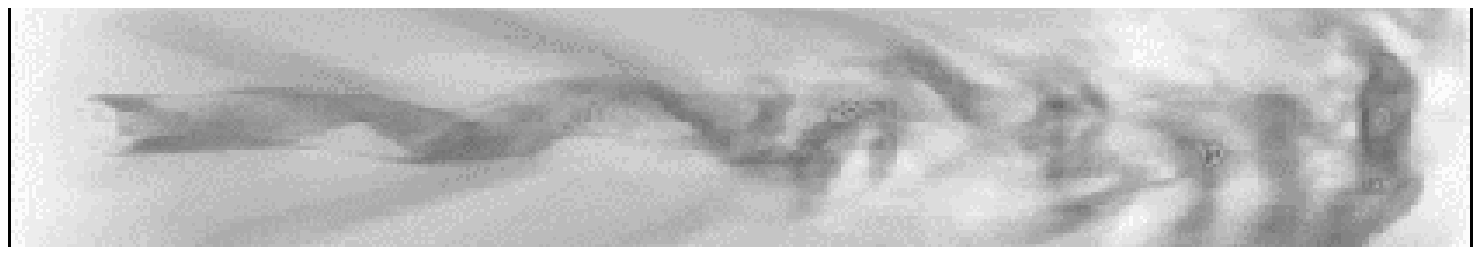}
\end{figure}

\noindent Fig. 5---Composite $\rho^2 T$ volumetric rendering of the Mach 5 DM
jet at $\omega R_{jt}/u=0.4$ shown in Figure 4.  The jet is inclined at
$45^o$ to the line of sight and the volume shown is $60R_{jt} \times
8R_{jt} \times 8R_{jt}$. Darker shading indicates higher values of
$\rho^2 T$.

\newpage

\begin{figure}[h!]
\vspace{10cm}
\includegraphics{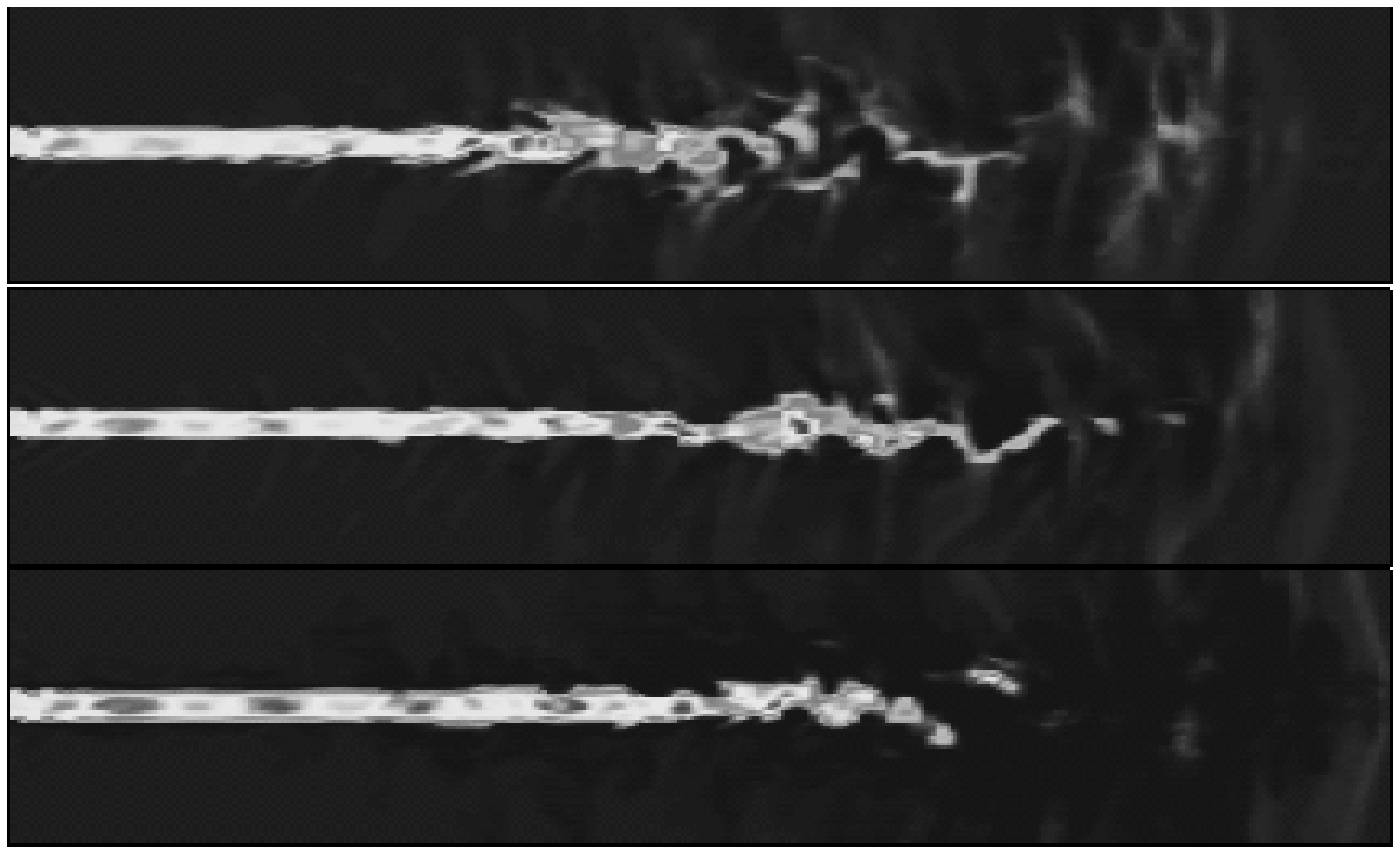}
\end{figure}

\noindent Fig. 6---Center cuts in density of Mach 5 AD (bottom), DM (middle),
and MB (top) jets at $\omega R_{jt}/u=1.0$.  Mapping to greyscale is
the same in each panel and the panel size is $300R_{jt} \times
20R_{jt}$ and underscaled horizontally by a factor of 3. In general,
lighter shading indicates higher density but in the jet the darkest
shading indicates the highest densities. Densities range from $\approx
60$~cm$^{-3}$ in the ambient to a maximum of $\approx 1,200$~cm$^{-3}$
in the jet.

\newpage

\begin{figure}[h!]
\vspace{16cm}
\includegraphics{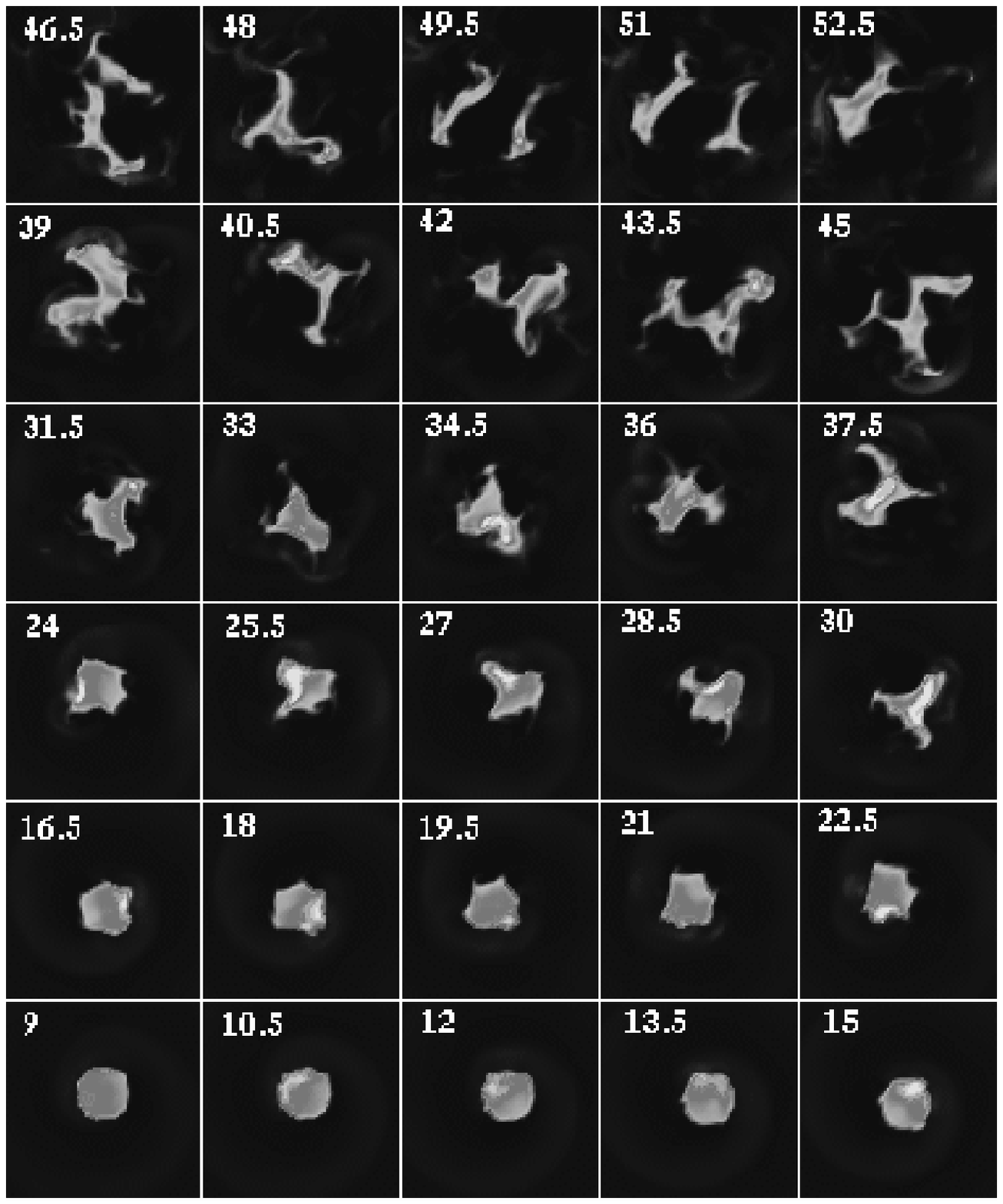}
\end{figure}

\noindent Fig. 7---Mach 5 DM jet cross sections at $\omega R_{jt}/u=0.4$.  The
number in each frame indicates the distance from the inlet in units of
R$_{jt}$. In general,  a lighter shade in these cross sections
indicates a higher density but in the jet the darkest shading within
lighter regions indicates the highest densities. Densities range from
$\approx 60$~cm$^{-3}$ in the ambient to a maximum of $\approx
1,200$~cm$^{-3}$ in the jet.

\newpage

\begin{figure}[h!]
\vspace{6cm}
\includegraphics{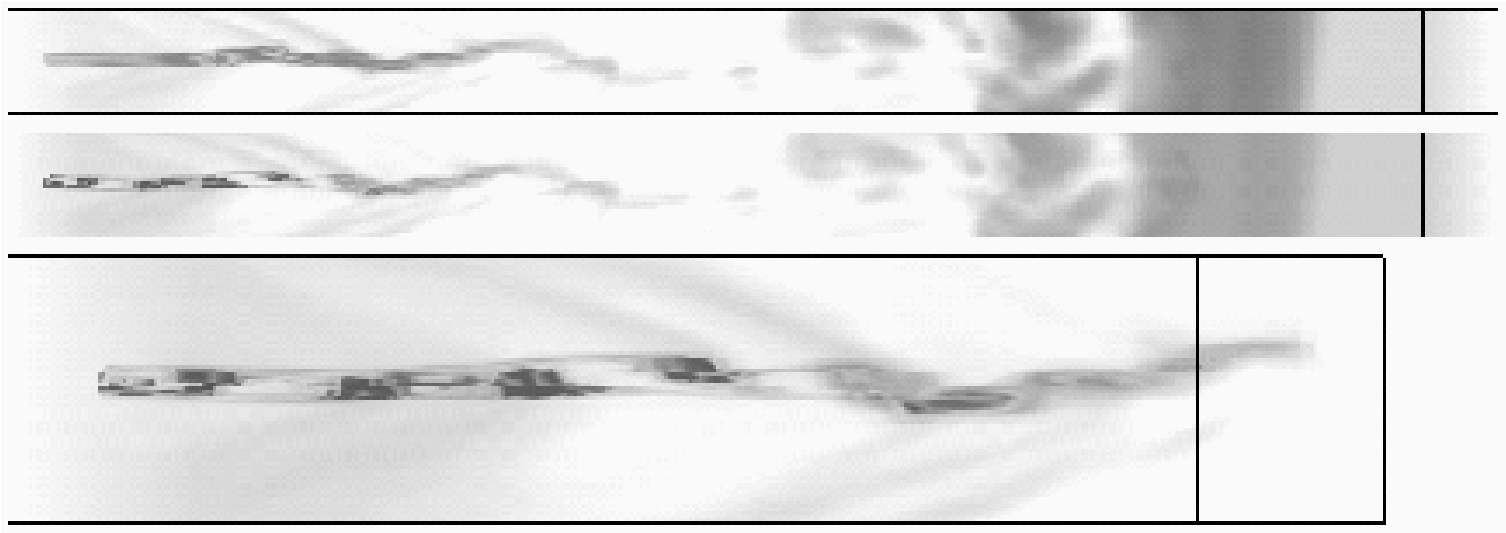}
\end{figure}

\noindent Fig. 8---Composite $\rho^2 T$ volumetric rendering of the Mach 5 DM
jet at $\omega R_{jt}/u=0.1$.  The jet is inclined at $45^o$ to the
line of sight. Panels show the entire jet (top), and entire jet
excluding material with $T<1000$~K (middle) with volume $300R_{jt}
\times 16R_{jt} \times 16R_{jt}$. Lower panel shows the knot region
excluding material with $T<1000$~K with volume $50R_{jt} \times
16R_{jt} \times 16R_{jt}$. Darker shading indicates higher values of
$\rho^2 T$ and reveals the hot dense parts of the jet.

\newpage

\begin{figure}[h!]
\vspace{6.8cm}
\includegraphics{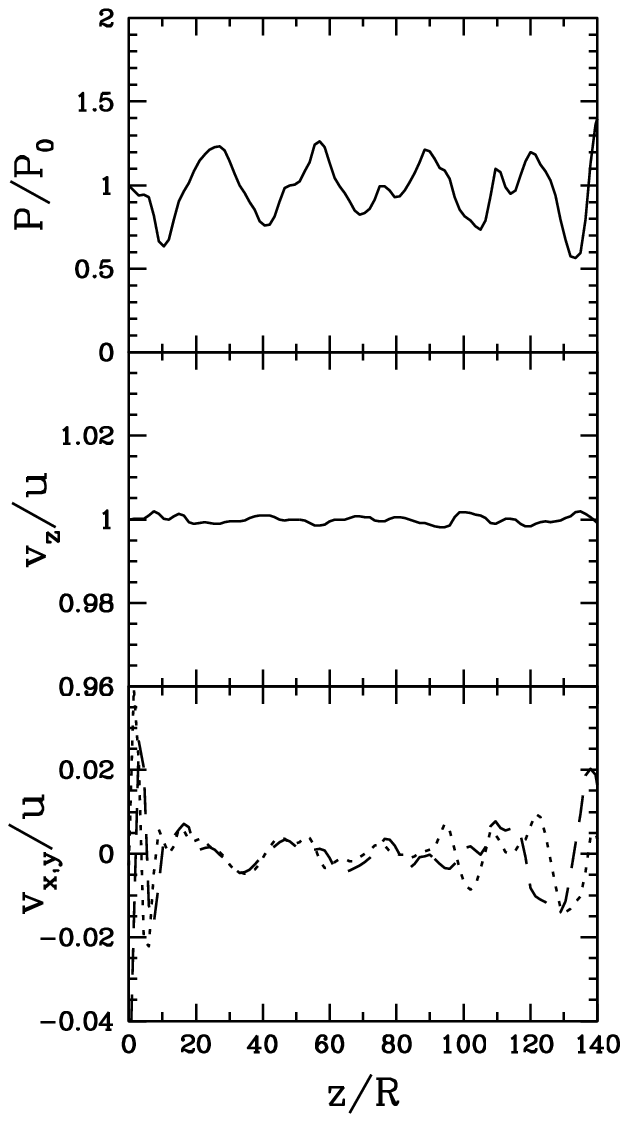}
\end{figure}

\noindent Fig. 9---Plots of pressure (top), axial velocity (middle) and transverse velocity components (bottom) along the jet axis from the Mach 5 DM jet at $\omega R_{jt}/u=1.0$.

\begin{figure}[h!]
\vspace{9.9cm}
\includegraphics{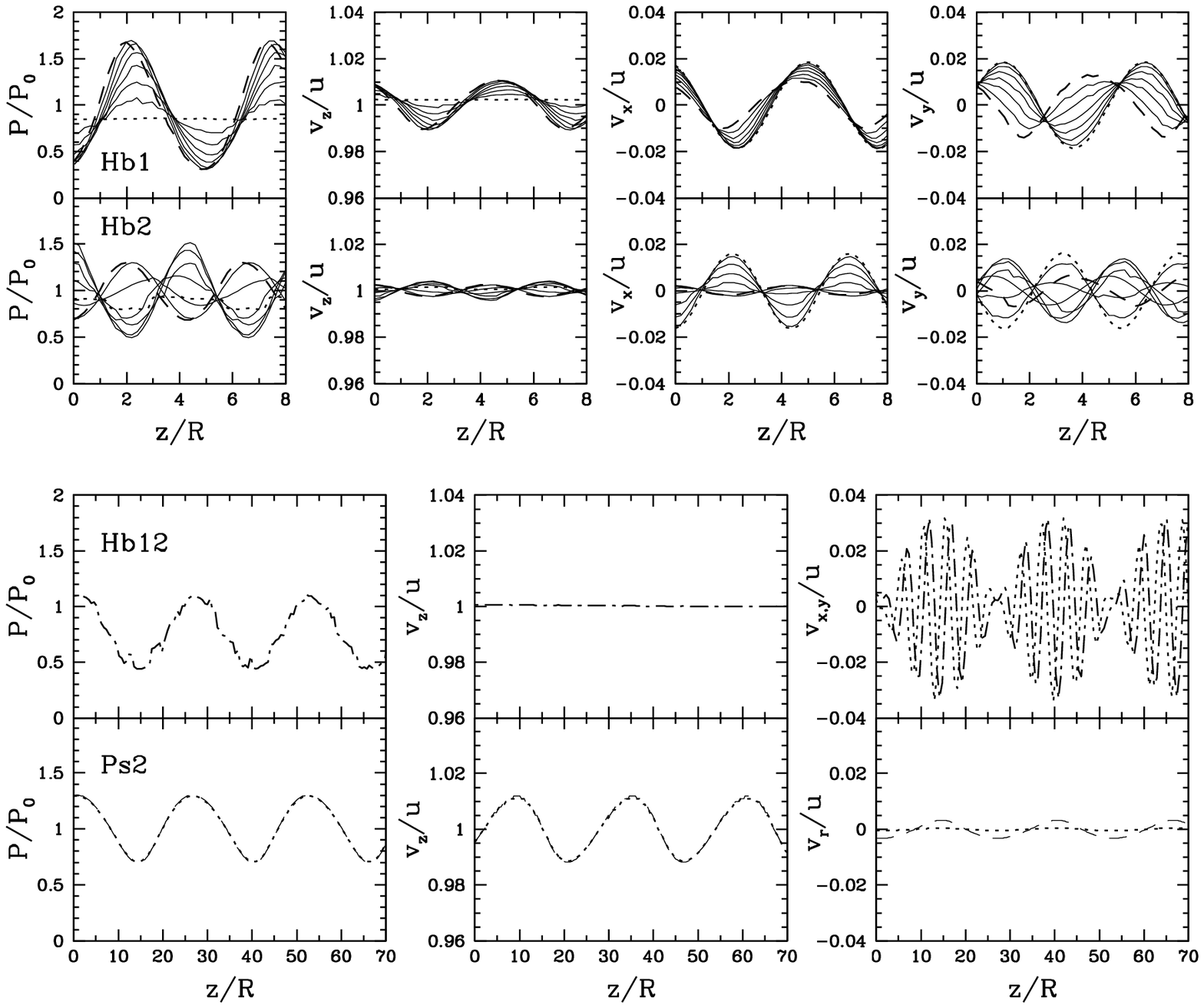}
\end{figure}

\noindent Fig. 10---(top two rows) Pressure, axial velocity ($v_z$), azimuthal
velocity ($v_x$) and radial velocity ($v_y$) for the first two helical body modes Hb1 and
Hb2 on a Mach 5 DM jet along 1D cuts parallel to the $z$-axis at radial
locations $r/R=0/8,1/8,2/8,...,7/8$ on the $+y$-axis. The outermost
(innermost) radial locations are indicated by the dashed (dotted)
lines. (bottom two rows) Pressure, axial velocity ($v_z$) and
transverse velocity components for wave-wave interaction between the
two helical body modes (Hb12) along the jet axis, and for the pinch cooling
mode (Ps2).  For Ps2, $v_r$ is shown on the axis (dotted line) and near
to the jet surface (dashed line).

\newpage

\begin{figure}[h!]
\vspace{8cm}
\includegraphics{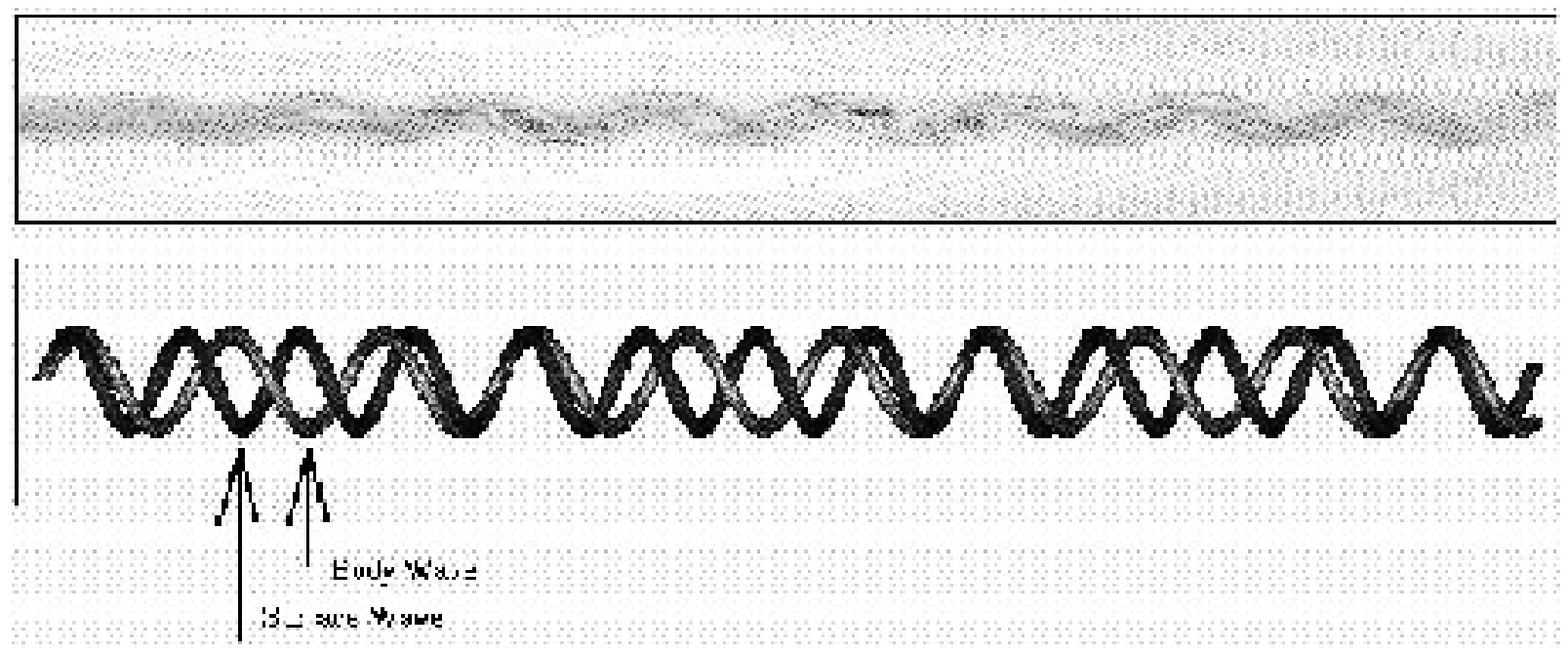}
\end{figure}

\noindent Fig. 11---Composite $\rho^2 T$ volumetric rendering of the Mach 20 DM
jet at $\omega R_{jt}/u=1.0$ (top) and illustration of the interaction
between surface and body wave (bottom).  The size shown is $60R_{jt}
\times 8R_{jt}$. Darker shading in the volumetric rendering indicates higher
values of $\rho^2 T$.

\newpage

\begin{figure}[h!]
\vspace{6.5cm}
\includegraphics{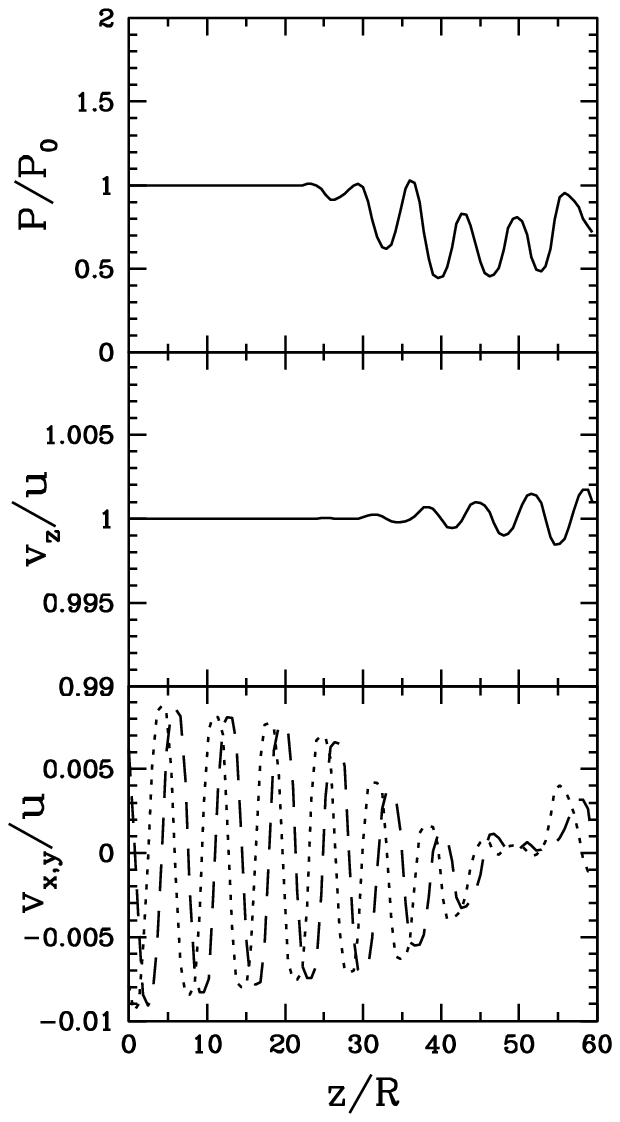}
\end{figure}
\noindent Fig. 12---Plots of pressure (top), axial velocity (middle) and transverse velocity components (bottom) along the jet axis from the Mach 20 DM jet at $\omega R_{jt}/u=1.0$.

\begin{figure}[h!]
\vspace{10.7cm}
\includegraphics{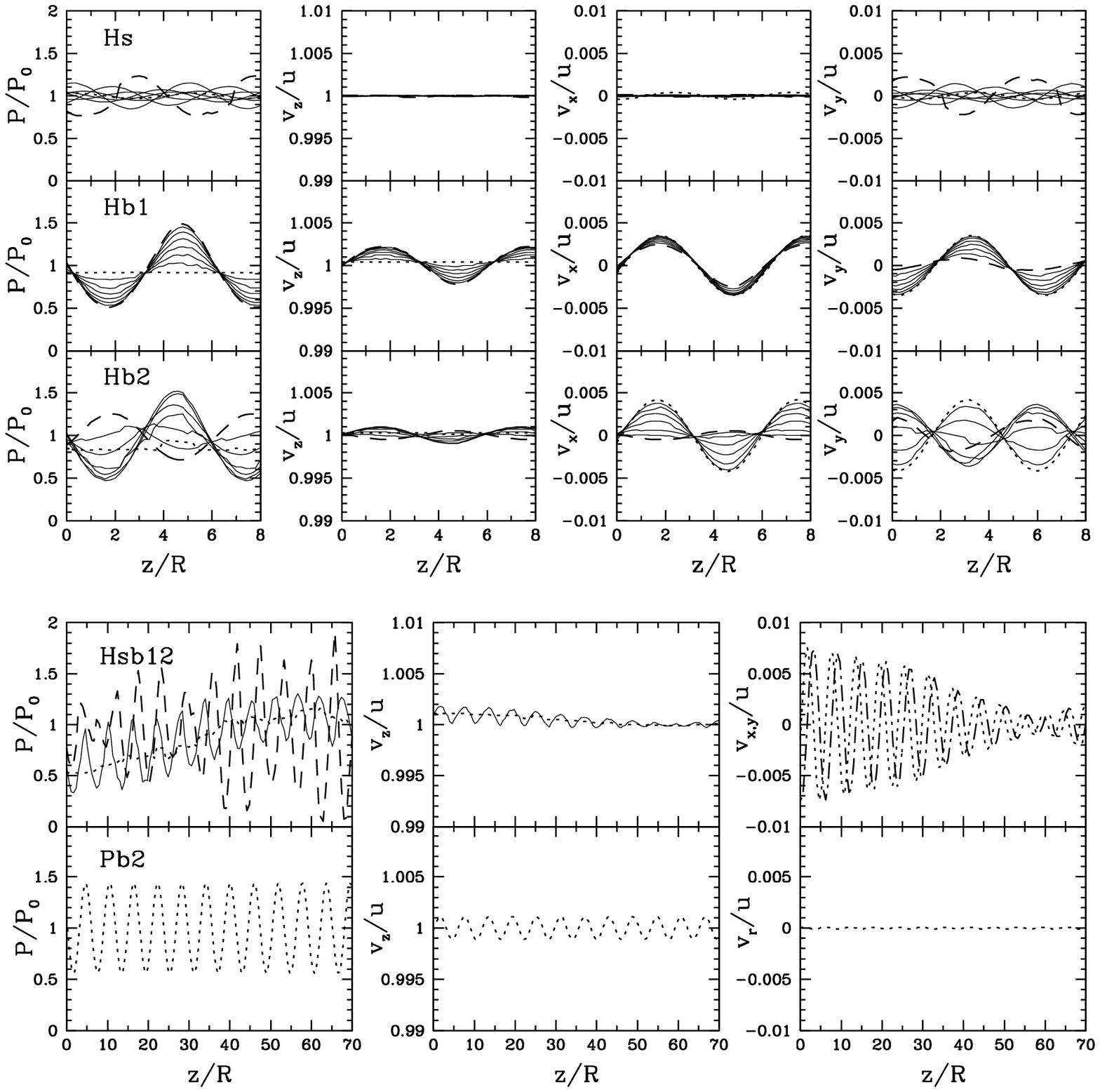}
\end{figure}

\noindent Fig. 13---Similar to Figure 10 but for helical surface (Hs), helical
body modes Hb1 and Hb2 and pinch body mode Pb2 on a Mach 20 DM jet. The
pressure panel for the combined helical surface and body waves (Hsb12)
shows 1D cuts along the jet axis (dotted line), at r/R = 1/8 (solid
line) and at r/R = 7/8 (dashed line). The axial velocity panel shows 1D
cuts along the jet axis (dotted line), at r/R = 1/8 (solid line).
Pressure, axial velocity ($v_z$) and radial velocity for the pinch body
mode (Pb2) are plotted along the jet axis in the lower panels.

\end{document}